# Superhalogen and Superacid

**Andrey V. Kulsha,**[1] **Dmitry I. Sharapa**[2,3]

Correspondence to: Andrey V. Kulsha Andrey_601@tut.by; Dmitry I. Sharapa dmitry.sharapa@kit.edu

[1] *Andrey V. Kulsha*
*Lyceum of Belarusian State University, 8 Ulijanauskaja str., Minsk, Belarus, 220030*
[2] *Dmitry I. Sharapa*
*Chair of Theoretical Chemistry and Interdisciplinary Center for Molecular Materials, Friedrich-Alexander-Universitat Erlangen-Nürnberg, Egerlandstraße 3, 91058 Erlangen, Germany*
[3] *Institute of Catalysis Research and Technology (IKFT), Hermann-von-Helmholtz-Platz 1, Eggenstein-Leopoldshafen, Germany, D-76344*

**ABSTRACT**

A superhalogen $F@C_{20}(CN)_{20}$ and a corresponding Brønsted superacid were designed and investigated on DFT and DLPNO-CCSD(T) levels of theory. Calculated compounds have outstanding electron affinity and deprotonation energy, respectively. We consider superacid $H[F@C_{20}(CN)_{20}]$ to be able to protonate molecular nitrogen. The stability of these structures is discussed, while some of the previous predictions concerning neutral Brønsted superacids of record strength are doubted.

## Introduction

Traditionally, a superhalogen is a molecule with high electron affinity, which forms a stable anion. A good example is $AuF_6$ with electron affinity of about 8.2 eV.[1] Superhalogen anions usually have low proton affinities leading to superacids. [2,3]

However, few of the known superhalogen neutral molecules are stable in condensed phase. For example, some anions were predicted to have vertical electron detachment energies above 13 eV,[4,5] but the corresponding neutral molecules are too unstable to be qualified as superhalogens. The goal of this article was to design structures that could behave closer to real halogen atoms and acid molecules, but with extreme properties.

### *Design*

The idea of symmetric cage surrounded with electron-withdrawing groups is barely new.[6-13] For example, one of the strongest currently known Brønsted acids is fluorocarborane acid $H[CHB_{11}F_{11}]$.[14] Its anion is formed by a carborane cage surrounded with fluorine atoms and shows superhalogenic behavior. A similar structure with cyano groups, $B_{12}(CN)_{12}^{2-}$, was suggested as a highly stable dianion with second electron bound by 5.3 eV .[15,16]

Going this way, we designed a dodecahedrane cage with 20 cyano groups. A large noble-gas-like HOMO-LUMO gap should make this structure kinetically more stable than halogenated dodecahedranes like $C_{20}Cl_{20}$ suffering from lone pair crowding and consequent elimination of halogen diatomics.[17] Therefore, the last step to superhalogen was the encapsulation of a single fluorine atom. It is known that atomic fluorine has lower electron affinity than atomic chlorine because of large density of negative charge in fluoride anion due to its small radius. Encapsulation into $C_{20}(CN)_{20}$ helps to delocalize the excess of negative charge over twenty electron-withdrawing groups and makes $F@C_{20}(CN)_{20}^{-}$ a very stable anion (Figure 1). Hereinafter we abbreviate $F@C_{20}(CN)_{19}$ structure as **X**, so the short notation for the superhalogenic anion is NC**X**⁻.

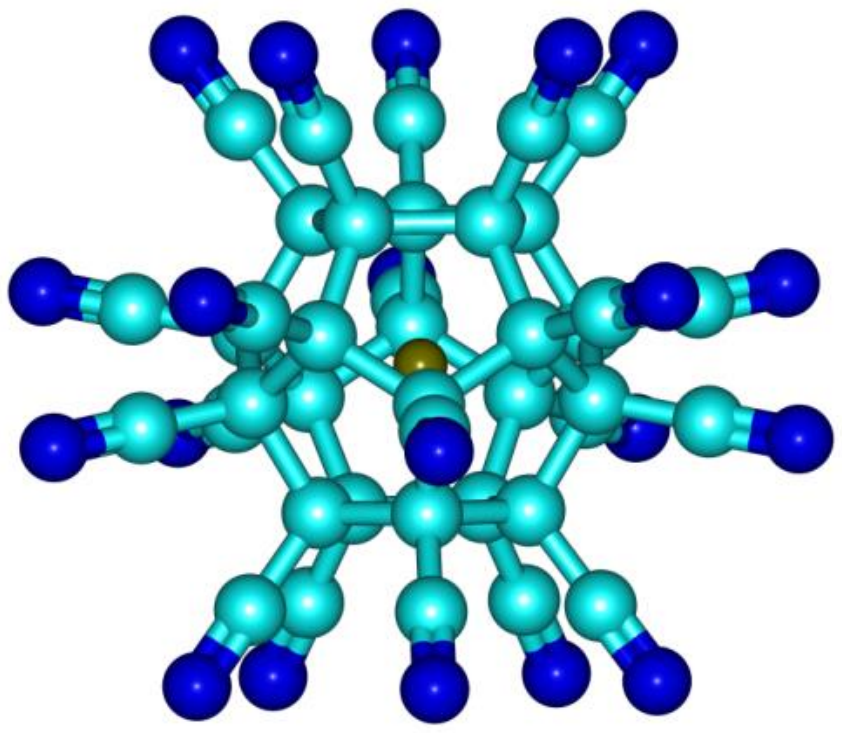

Figure 1. Structure of $F@C_{20}(CN)_{20}^{-}$.

Color coding: C – cyan, N – dark blue, F – olive.

### *Computational details*

Recently developed DLPNO approach to coupled cluster theory[18-23] induced us to use this level of theory for the single-point computations as implemented in ORCA 4.0 package.[24] This method was shown to provide results of canonical coupled-cluster quality.[25-31] All the calculations were performed for gas phase.

The standard cc-pVTZ basis set[32] was chosen for all atoms except for nitrogen, where maug-cc-pVTZ[33] was used because nitrogen atoms fill the surface of the anion. The corresponding auxiliary basis sets were cc-pVTZ/C and aug-cc-pVTZ/C,[34] respectively. Test calculations (Table 1) were performed with 6-311G* basis set.[35]

For geometry optimizations we chose a DFT functional PBE0[36] as implemented in Gaussian16 package[37]. This functional is known to produce overall good geometries for organic structures. Usually it underestimates C≡N bond lengths nearly by 0.01Å, and this is the most energy-sensitive deviation for our systems, but the corresponding shift of energy differences appeared to be negligible for our purposes (about 0.002 eV). Vibrational frequencies were computed on PBE0 level of theory to ensure that the structures being discussed correspond to real minima on their PES. To obtain reaction enthalpies (including electron affinity EA and proton affinity PA) at 0 Kelvins, zero-point vibrational energies (ZPE) were accounted. The corresponding entropies and thermal corrections were used to estimate standard Gibbs energies (including gas-phase basicity GB) at 298.15 Kelvins.

Most of the structures accounted are closed-shell singlets with HOMO-LUMO gap sufficiently large to use NormalPNO cutoffs in ORCA. However, neutral NC**X** is a doublet with quite small HOMO-LUMO gap; hence we used UHF wavefunction with TightPNO cutoffs for electron affinity determination. [26,27,38] That choice is confirmed with results of test computations collected in Table 1.

Table 1. Proton affinity of NC**X**⁻ and electron affinity of NC**X**. NormalPNO vs. TightPNO approximations. PBE0 geometry, 6-311G* basis set, no ZPE corrections. (Test calculations)

| Method | PA(NC**X**⁻) eV | EA(NC**X**) eV |
|---|---|---|
| Canonic CCSD [a] | 9.438 | 11.131 |
| DLPNO-CCSD / NormalPNO | 9.433 | 10.909 |
| DLPNO-CCSD / TightPNO | 9.438 | 11.157 |
| DLPNO-CCSD(T) / NormalPNO | 9.436 | 10.538 |
| DLPNO-CCSD(T) / TightPNO | 9.444 | 10.649 |

[a] performed in Gaussian16.

### *Electron affinity*

The ground state of NC**X**⁻ anion is a closed-shell singlet of $I_h$ symmetry. According to PBE0 functional, cage C—C bond lengths are 1.586Å, so the distance from the center fluorine to cage carbon is 2.222Å. Outer C—C bond lengths are shortened to 1.460Å, while C≡N bond lengths are 1.148Å.

Neutral NC**X** radical has an open-shell doublet ground state with lower symmetry $D_{3d}$ due to Jan-Teller effect. However, the stretch along $C_3$ axis is quite small (only +0.008Å between axial atoms of the cage), so it may be smoothed by zero-point vibrations. With almost zero spin density on the central fluorine, neutral NC**X** looks more like fluoride anion inserted in positively charged cage instead of neutral fluorine atom in neutral cage. Table 2 shows the results of electron affinity calculations.

Table 2. Electron affinity of NC**X**.

| Source | EA(NC**X**) / eV |
|---|---|
| PBE0 | 9.766 |
| PBE0 ZPE correction | –0.161 |
| DLPNO-CCSD | 11.428 |
| DLPNO-CCSD(T) | 10.913 |
| DLPNO-CCSD(T) + ZPE | 10.752 |

ZPE scaling, basis set enrichment and quadruplets accounting should probably push the final value a bit higher, so we round our result to 10.8 eV. That makes NC**X**, to our knowledge, a superhalogen with the highest electron affinity ever predicted. However, higher affinities should be possible, and an example will be shown later in section „Possible improvements".

***Superacid***

The most favorable protonation site in NC**X**$^-$ is nitrogen atom. The resulting molecule HNC**X** has singlet ground state of $C_{3v}$ symmetry with dipole moment of 13.96 Debye and NH bond length of 1.008Å according to PBE0 functional. Proton affinity of NC**X**$^-$ was estimated as 9.3 eV (Table 3), corresponding to gas-phase basicity of 208 kcal/mol.

Table 3. Proton affinity of NC**X**$^-$.

| Source | PA(NC**X**$^-$) / eV |
|---|---|
| PBE0 | 9.681 |
| PBE0 ZPE correction | –0.276 |
| DLPNO-CCSD | 9.586 |
| DLPNO-CCSD(T) | 9.579 |
| DLPNO-CCSD(T) + ZPE | 9.303 |

HNC**X** superacid also features the strongest single bond known. Indeed, the dissociation energy HNC**X** = H + NC**X** could be estimated as PA(NC**X**$^-$) + EA(NC**X**) – IE(H) = 9.3 eV + 10.8 eV – 13.6 eV = 6.5 eV, or about 150 kcal/mol.

We decided to compare the acidity of HNC**X** with the acidity of fluorocarborane superacid $H[CHB_{11}F_{11}]$ which is able to protonate $CO_2$.[39] It is known that in highly acidic media proton tends to form bridged structures,[40] so we compared both HA and AHA$^-$ for A = $CHB_{11}F_{11}$ and A = NC**X**.

The most favorable protonation site in $CHB_{11}F_{11}$ appeared to be fluorine atom most distant from carbon, leading to $C_s$ symmetry for neutral $H[CHB_{11}F_{11}]$ and $C_{2h}$ symmetry for $H[CHB_{11}F_{11}]_2^-$ anion (Figure 2), while H[NC**X**]$_2^-$ anion features $D_{3d}$ symmetry. Both bridged anions contain short symmetrical hydrogen bonds with N···H distances of 1.259Å in H[NC**X**]$_2^-$ and F···H distances of 1.138Å in $H[CHB_{11}F_{11}]_2^-$.

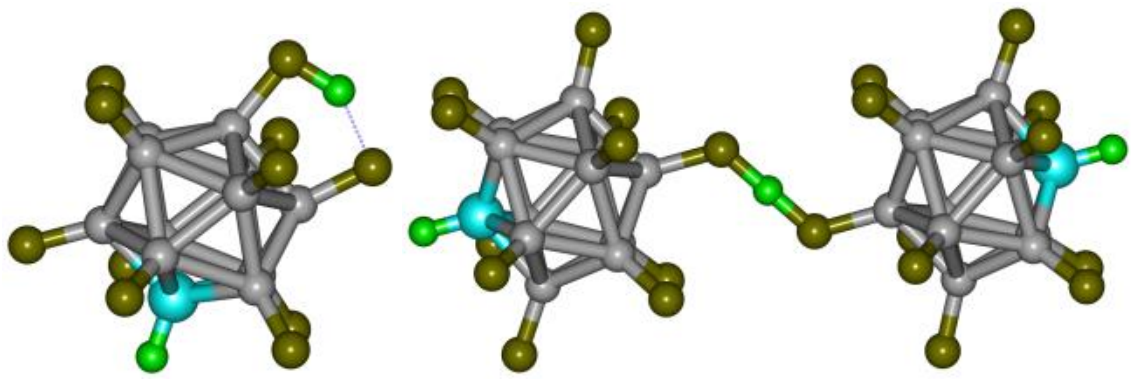

Figure 2. $H[CHB_{11}F_{11}]$ (left) and $H[CHB_{11}F_{11}]_2^-$ anion (right). Color coding: C – cyan, B – silver, F – olive, H – green.

Table 4. Acidity comparison of HNC**X** and $H[CHB_{11}F_{11}]$. DLPNO-CCSD(T) + PBE0 thermal correction (kcal/mol).

| Proton binding way | A = NC**X** | A = $CHB_{11}F_{11}$ [a] |
|---|---|---|
| $H^+ + A^- = HA$ | –208.1 | –221.0 |
| $H^+ + 2A^- = AHA^-$ | –229.3 | –249.6 |

[a] maug-cc-pVTZ basis set was used for fluorine atoms.

According to gas-phase protonation Gibbs energies shown in Table 4, HNC**X** should be much stronger acid than $H[CHB_{11}F_{11}]$, so we wondered the capability of HNC**X** to protonate molecular nitrogen and to produce a bridged cation $[NNHNN]^+$ first observed in 1999.[41]

We considered the ion pair $[NNHNN]^+$[NC**X**]$^-$ and compared it to similar ion pairs formed by 1,2-dicyanocyclopentadiene and pentacyanotoluene in acetonitrile solution (Table 5). For $[NNHNN]^+$[NC**X**]$^-$ we found the ground minimum geometry of $C_s$ symmetry with almost linear NNHNN trapped between three nearby CN needles. $[(CH_3CN)_2H]^+[C_5H_3(CN)_2]^-$ appeared to have $C_{2v}$ symmetry with symmetrical cation embowed around two cyano groups of anion, while in $[(CH_3CN)_2H]^+[CH_2C_6(CN)_5]^-$ methyl groups of cation snuggle to cyano groups in *ortho*- and *para*-positions of benzene ring (Figure 3).

Table 5. Proton solvation comparison in acetonitrile and nitrogen. DLPNO-CCSD(T) + PBE0 thermal correction.

| Gas phase reaction | ΔG (kcal/mol) |
|---|---|
| HNC**X** + $2N_2$ = $[NNHNN]^+[NC\mathbf{X}]^-$ | 38.4 |
| $H[C_5H_3(CN)_2]$ + $2CH_3CN$ = $[(CH_3CN)_2H]^+[C_5H_3(CN)_2]^-$ | 37.8 |
| $H[CH_2C_6(CN)_5]$ + $2CH_3CN$ = $[(CH_3CN)_2H]^+[CH_2C_6(CN)_5]^-$ | 50.0 |

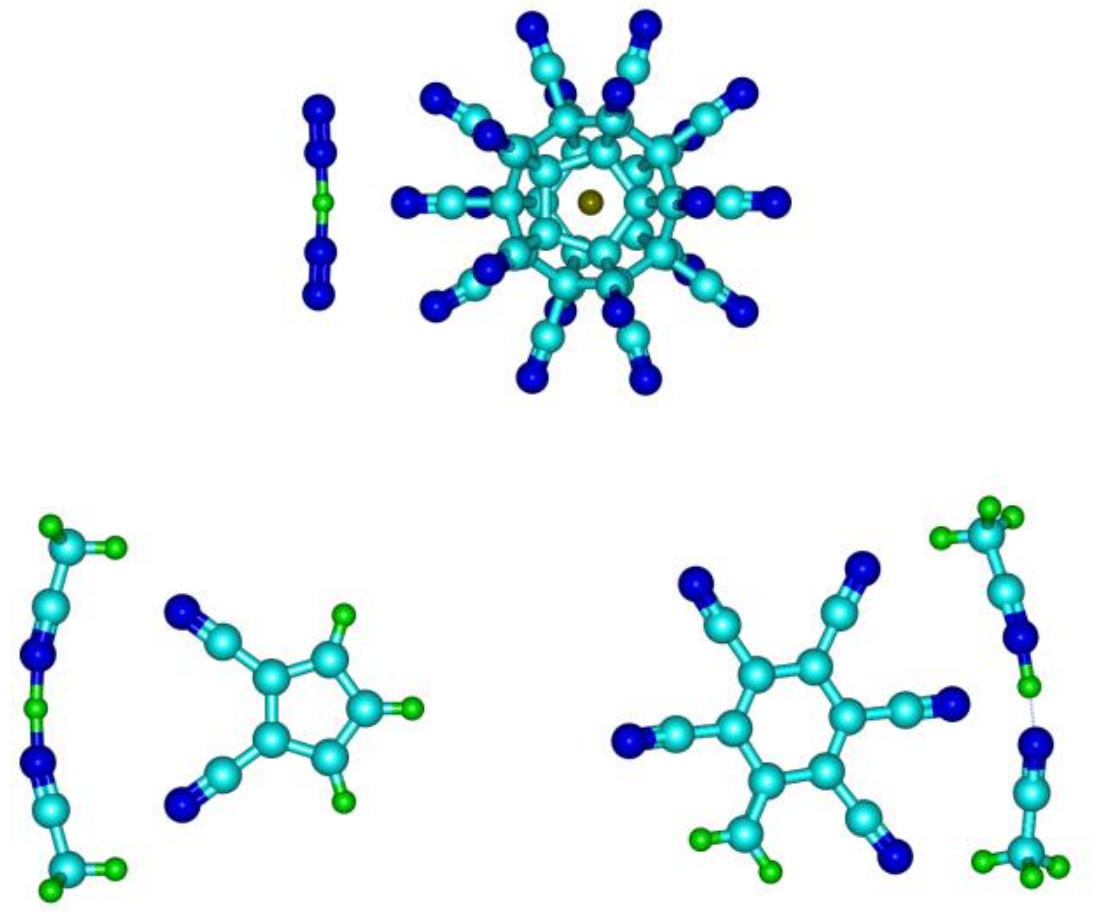

Figure 3. $[NNHNN]^+[NC\mathbf{X}]^-$ (top), $[(CH_3CN)_2H]^+[C_5H_3(CN)_2]^-$ (bottom left), $[(CH_3CN)_2H]^+[CH_2C_6(CN)_5]^-$ (bottom right). Color coding: C – cyan, N – dark blue, H – green.

Experimental pKa values in $CH_3CN$ are 10.17 for 1,2-dicyanocyclopentadiene[42] and 20.14 for pentacyanotoluene[43]; so simple linear interpolation gives us pKa = 10.64 in nitrogen for HNC**X**. However, that very rough result should be treated only as qualitative estimation providing some evidence that molecular nitrogen could be indeed protonated by HNC**X** superacid.

The possibility of auto-ionization of HNC**X** was also considered. Neutral acid has five possible protonation sites to form HNC**X**H$^+$ tautomers. Their energy depends on the distance between protons (the larger is distance, the lower is energy), so the most favorable tautomer has $D_{3d}$ symmetry. That fact suggests that in condensed phase HNC**X** will probably form a linear polymer ···H···[NC**X**]···H···[NC**X**]··· instead of cyclic or zigzag-shaped clusters typical for HF. The calculated gas-phase basicity of neutral HNC**X** is 164 kcal/mol; accounting earlier results, the Gibbs energy of the autoionization process 3HNC**X** = $H[NC\mathbf{X}]_2^-$ + HNC**X**H$^+$ appeared to be as small as 23 kcal/mol in gas phase (Table 6), so we expect a high degree of autoionization for HNC**X** in a solution.

Table 6. Autoionization energy for HNC**X**. DLPNO-CCSD(T) + PBE0 thermal correction.

| Gas phase reaction | ΔG (kcal/mol) |
|---|---|
| HNC**X** = NC**X**$^-$ + H$^+$ | 208.1 |
| HNC**X** + H$^+$ = HNC**X**H$^+$ | –164.2 |
| HNC**X** + NC**X**$^-$ = $H[NC\mathbf{X}]_2^-$ | –21.2 |
| 3HNC**X** = $H[NC\mathbf{X}]_2^-$ + HNC**X**H$^+$ | **22.7** |

$SbF_5$ could be a proper solvent for HNC**X** because of acidity-enhancing N···Sb coordination and quite low basicity of $SbF_5$ itself.

***Other simple derivatives***

As well as common halogen atoms, neutral NC**X** readily forms covalent bonds with atoms other than hydrogen and can also exist as a dimer $[NC\mathbf{X}]_2$. Actual bonds are formed by nitrogen atoms of cyano groups.

The $[NC\mathbf{X}]_2$ molecule is analogous to halogen diatomics (Figure 4). It has singlet ground state of $D_{3d}$ symmetry featuring NN bond of length 1.238Å which is close to double bond in $N_2H_2$. Another molecule, FNC**X**, is an analog of interhalogen diatomics. It has $C_{3v}$ singlet ground state with dipole moment of 13.04 Debye that is almost as large as of HNC**X**, suggesting an outstanding electronegativity of our superhalogen. Quite short FN bond of 1.249Å shows partially double character just like FC bond in FCN.

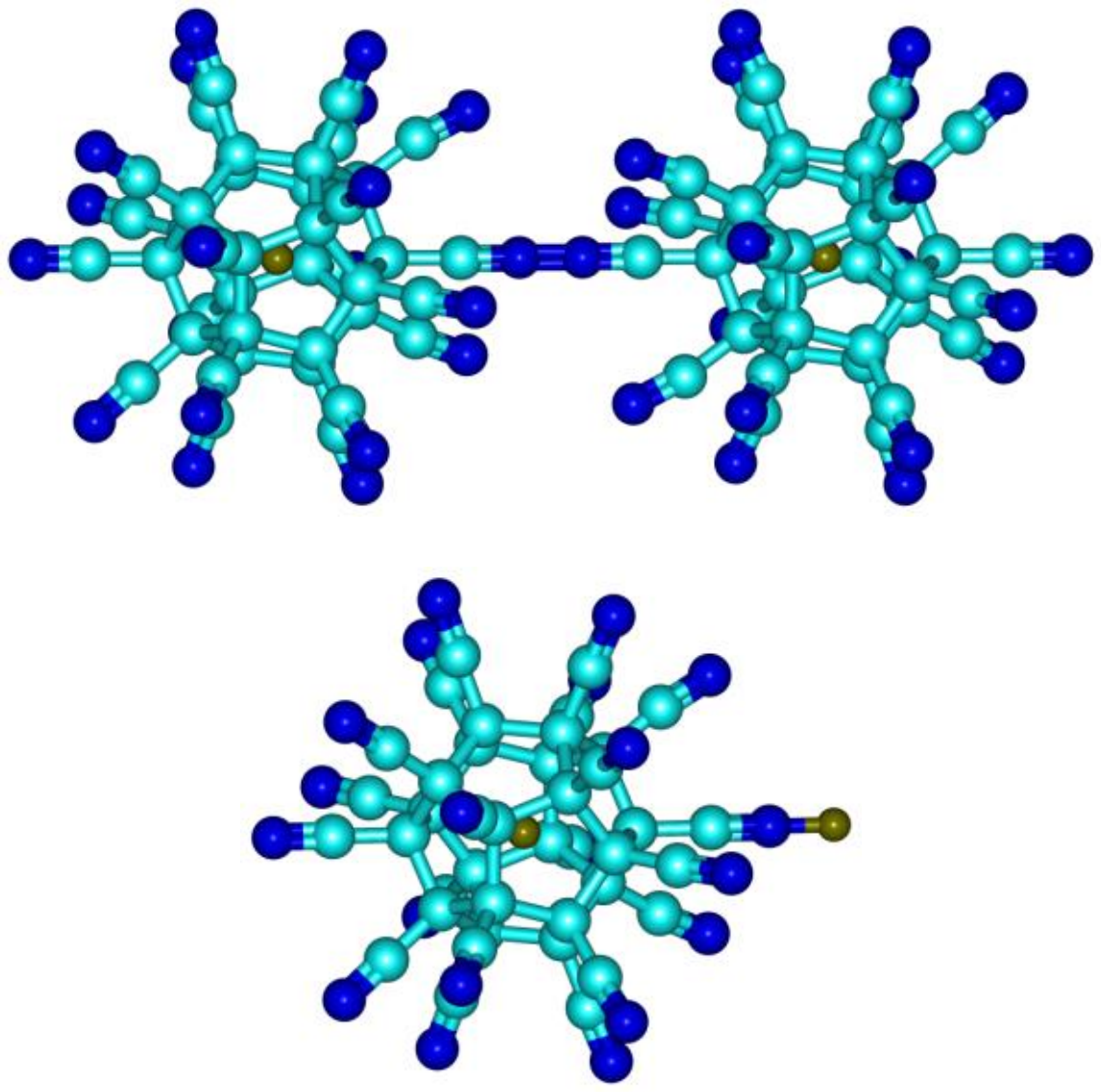

Figure 4. $[NC\mathbf{X}]_2$ (top) and $FNC\mathbf{X}$ (bottom). Color coding: C – cyan, N – dark blue, F – olive.

Direct calculation of dissociation energies (DE) for $[NC\mathbf{X}]_2$ and $FNC\mathbf{X}$ would need UHF wavefunction with TightPNO cutoffs. In order to avoid heavy calculations, we made sideway estimates using gas phase reactions between closed-shell structures (Table 7).

Table 7. Supplemental calculations for dissociation energy estimations. DLPNO-CCSD(T) + PBE0 ZPE correction.

| Gas phase reaction | Energy / eV |
|---|---|
| $[NC\mathbf{X}]_2 + 2e^- = 2NC\mathbf{X}^-$ | –17.439 |
| $[NC\mathbf{X}]_2 + HF = HNC\mathbf{X} + FNC\mathbf{X}$ [a] | –0.097 |

[a] Negative ΔH suggests that $[NC\mathbf{X}]_2$ would oxidize HF.

Thus, DE($[NC\mathbf{X}]_2$) was estimated as 2 · EA($NC\mathbf{X}$) – 17.439 eV = 4.065 eV ≈ 4.1 eV, or about 94 kcal/mol, which is much higher than for diatomic halogens. Taking into account previously estimated DE($HNC\mathbf{X}$) of 6.5 eV and experimental value of DE(HF) = 5.9 eV [44], we obtain DE($FNC\mathbf{X}$) = (4.1 + 5.9 + 0.1 – 6.5) eV = 3.6 eV, or about 83 kcal/mol.

***Stability***

It is clear that our predictions have little sense until the stability of our superhalogen and its derivatives is examined. First, the cage C—C bond in $NC\mathbf{X}^-$ is predicted to be about 0.05Å longer than C—C bond in ethane, hinting at some strain. Therefore, we wondered how much energy the encapsulation takes.

The encapsulation of fluorine atom, fluoride anion and neon atom (Table 8) was considered. Again, to avoid heavy calculations, the sideway reaction $C_{20}(CN)_{20}$ + HF = $HNC\mathbf{X}$ was used together with known values of EA(F) = 3.401 eV [45] and PA($F^-$) = 16.063 eV[44].

Table 8. Encapsulation energies for $F^-$, F and Ne. DLPNO-CCSD(T) + PBE0 ZPE correction.

| Gas phase reaction | Energy / eV |
|---|---|
| $C_{20}(CN)_{20}$ + Ne = Ne@$C_{20}(CN)_{20}$ | 3.055 |
| $C_{20}(CN)_{20}$ + HF = $HNC\mathbf{X}$ | 1.473 |
| $C_{20}(CN)_{20} + F^- = NC\mathbf{X}^-$ | –5.287 [a] |
| $C_{20}(CN)_{20}$ + F = $NC\mathbf{X}$ | 2.064 [b] |

[a] Calculated as 1.473 eV – PA($F^-$) + PA($NC\mathbf{X}^-$).
[b] Calculated as –5.287 eV – EA(F) + EA($NC\mathbf{X}$).

Exothermical encapsulation of fluoride anion is expectative because of negative charge delocalization noted earlier. However, for isoelectronic process of neon encapsulation we have 8.3 eV larger energy, suggesting some repulsion between the noble electron cloud and the cage. Surprisingly, for fluorine atom the encapsulation is 1 eV less endothermic despite its larger radius compared to neon. That could be explained with proximity of energies, and therefore significant correlation, between fluorine-localized and cage-localized molecular orbitals. In other words, the fluorine atom fits the cage in some sense.

In terms of Gibbs energy, discussed superhalogen is thermodynamically unstable against the loss of fluoride atom by 54 kcal/mol. However, tight pentagonal holes of the cage provide a very high kinetic barrier for this decomposition route. Such a barrier for smaller helium atom was experimentally proven to be so high that helium endocomplex survive for weeks at room temperature.[46]

The search for a transition state pointed to the fact that another decomposition route is energetically favorable. It is the cleavage of an outer C—C bond despite its partially double character. The key is that fluorine atom may attack one of the cage carbons from inside, pulling it inward and leaving the CN group alone, just like in $S_N2$. The resulting molecule **X** (structurally related to "in-adamantane"[47]) has a singlet ground state of $C_{3v}$ symmetry with C—F bond length of 1.431Å (Figure 5). That large bond length is the result of carbon cage strain.

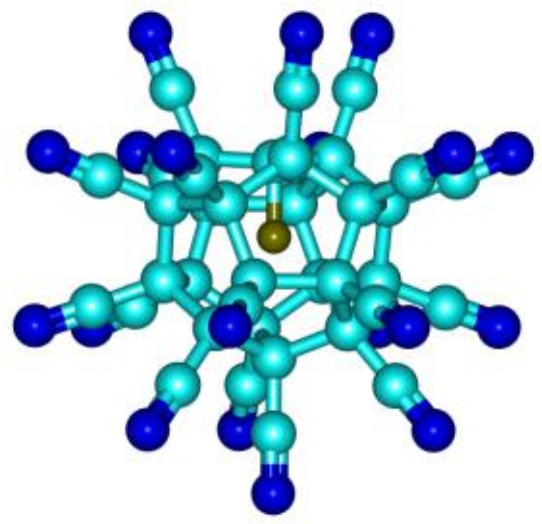

Figure 5. Structure of **X** molecule $C_{20}$(*in*-F)$(CN)_{19}$. Color coding: C – cyan, N – dark blue, F – olive.

The formation of **X** from superhalogen and its derivatives is considered in Table 9. As before, sideway calculations were performed using known values of EA(CN) = 3.862 eV and PA($CN^-$) = 15.199 eV[48].

Table 9. Stability against the formation of **X**. DLPNO-CCSD(T) + PBE0 ZPE correction.

| Gas phase reaction | Energy / eV |
|---|---|
| HNC**X** = HCN + **X** | 0.810 |
| NC**X**$^-$ = CN$^-$ + **X** | 6.707 [a] |
| NC**X** = CN + **X** | –0.183 [b] |

[a] Calculated as 0.810 eV + PA($CN^-$) – PA(NC**X**$^-$).
[b] Calculated as 6.707 eV + EA(CN) – EA(NC**X**).

These results show that the superhalogen anion is very stable against the cage concaving, while the superacid is moderately stable. The corresponding gas phase Gibbs energies were calculated to be +143.1 and +6.8 kcal/mol, respectively.

Considering kinetic stability on PBE0 level of theory, we found that neutral NC**X** might transform to $C_{20}$(*in*-F)$(CN)_{18}$(C(=N)CN) (Figure 6) through the transition state with distorted cyclobutane unit. The barrier of 23 kcal/mol suggests days of lifetime at 0°C, but we expect that DFT overestimates the kinetic stability of neutral superhalogenic radical, therefore somewhat lower temperatures will be needed to isolate it.

A similar transition state was also found for HNC**X**, but the calculated barrier of 45 kcal/mol makes the novel superacid stable up to 200°C.

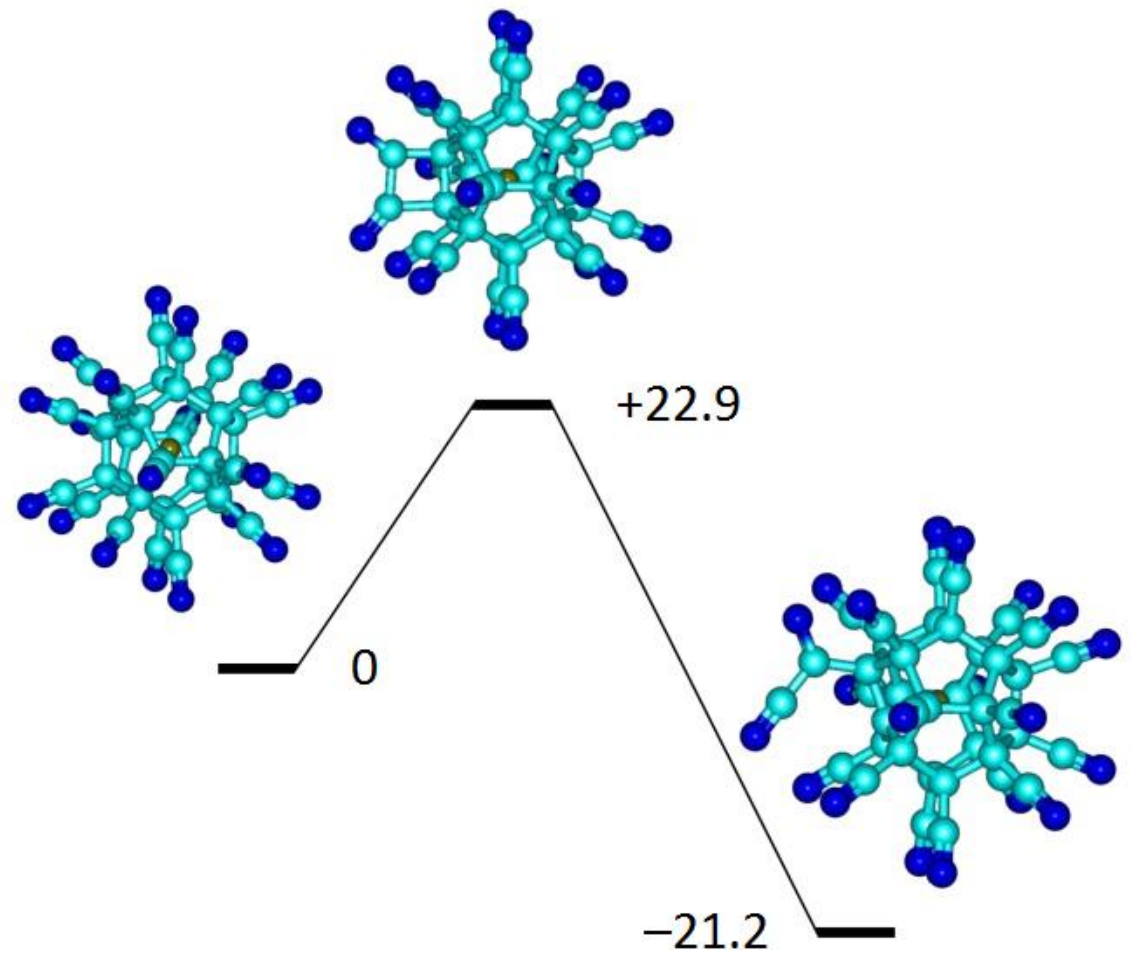

Figure 6. Intramolecular radical substitution reaction of NC**X**. PBE0 + thermal correction, kcal/mol.

### *Synthesis*

Despite a high amount of papers related to dodecahedrane endocomplexes,[49-55] there is only one up to date observed experimentally – He@$C_{20}H_{20}$, that was obtained by shooting a beam of helium ions at a continuously deposited surface of dodecahedrane.[46] Encapsulation was detected on the level of 0.01% that definitely cannot be accounted as a "preparative" technique. In our opinion, "molecular surgery" approach that was used for getting different endofullerene complexes, might be also applied in this case. This methodology involves inserting the “guest”-atom into the cavity of precursor that has reasonably big opening. Being non-experts in “molecular surgery”, authors guess that structures A and B (Figure 7) might be among possible intermediates. These structures are results of F-anion encapsulation into already known precursors of dodecahedrane.[56-58]

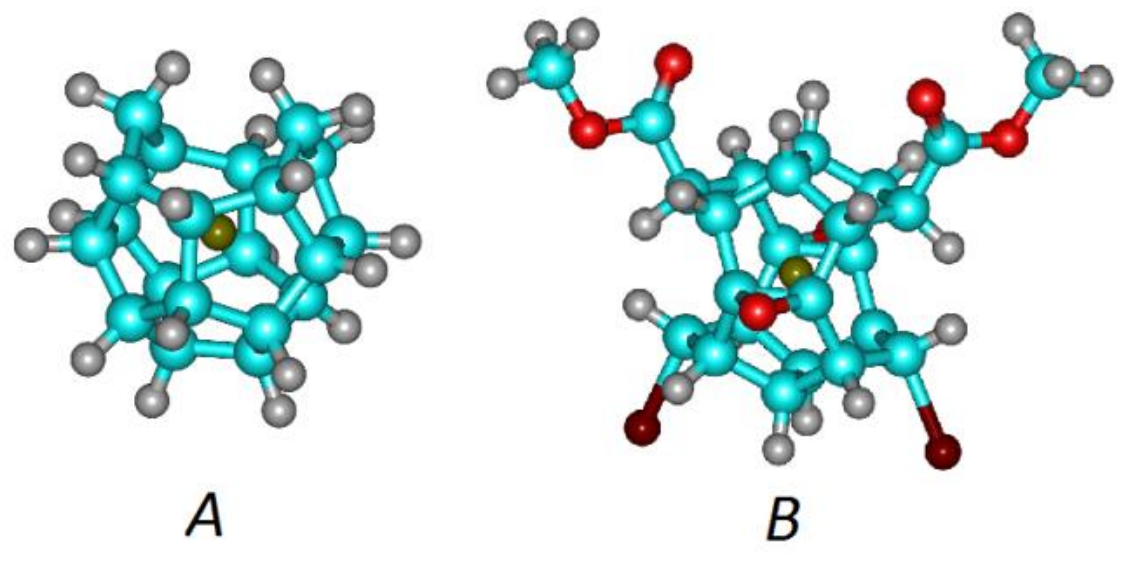

Figure 7. Possible intermediates for encapsulation of fluorine in dodecahedrane cage. Color coding: C – cyan, H – grey, O – red, Br – brown, F – olive.

Structure B seems to be especially attractive because: 1) according to Bertau,[56] further C—C coupling is performed with a fluoride salt of non-nucleophilic Schwesinger cation; 2) resulted dodecahedrane is highly functionalized that would simplify further transformation to NC**X**. Nevertheless, authors would like to note that the aim of current paper is mainly the demonstration of the concept, but not the detailed elaboration of synthetic conditions.

***Possible improvements***

The limits can always be pushed further. We compared our superhalogen to a few other structures, trying to improve either electron affinity or proton affinity value.

First of all, two larger carbon cages were considered: $C_{24}$ of $D_{6d}$ symmetry and $C_{28}$ of $T_d$ symmetry, both decorated with cyano groups (Figure 8). However, DFT estimates show that electron affinity does not grow this way (Table 10). Calculated Hartree-Fock HOMO energies for anions support this trend.

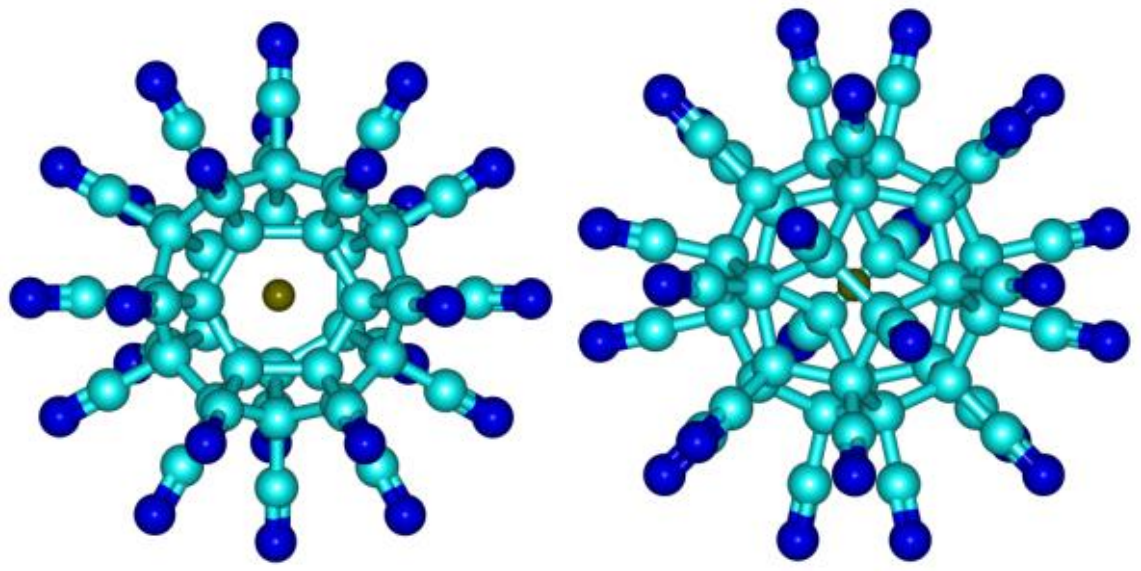

Figure 8. F@$D_{6d}$-$C_{24}(CN)_{24}$ (left) and F@$T_d$-$C_{28}(CN)_{28}$ (right). Color coding: C – cyan, N – dark blue, F – olive.

Table 10. Electron affinity comparison for different cages.

| Structure | EA / eV (PBE0 without ZPE corrections) | HOMO energy / eV (HF wavefunction for anion) |
|---|---|---|
| F@$C_{20}(CN)_{20}$ | 9.766 | –11.768 |
| F@$C_{24}(CN)_{24}$ | 9.502 | –11.482 |
| F@$C_{28}(CN)_{28}$ | 9.484 | –11.455 |

Thus, dodecahedrane cage seems to be the best fit for fluorine atom. Another try to improve the structure was changing cyano groups to $CF_3$ ones, but that quickly lead to overcrowding of fluorine atoms, preventing the proper delocalization of negative charge.

Looking at carborane cages, we noticed that the cage $CB_{11}$ of $C_{5v}$ symmetry, decorated with 12 $CF_3$ groups (Figure 9), has less crowding and was already proposed in 2000.[59] As well as $B_{12}(CF_3)_{12}$ and related structures, it was suggested by Ivo Leito and coworkers [60,61] as an anion for a Brønsted superacid of record-breaking strength. Unfortunately, we found that claim to be doubtful since $CF_3$ group is not stable in highly acidic media. To be precise, fluorine atom appears to be the most favorable protonation site, and once it is protonated, HF molecule is eliminated from the structure without any notable barrier. That happens because of larger electronegativity of carbon compared to hydrogen, which makes H—F single bond stronger than C—F single bond. The authors were aware of that issue.[62]

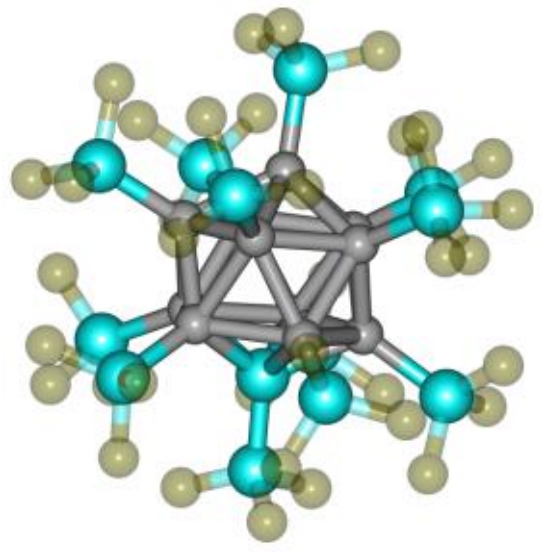

Figure 9. $CB_{11}(CF_3)_{12}^-$ anion. Color coding: C – cyan, B – silver, F – transparent olive.

As for electron affinity of $CB_{11}(CF_3)_{12}$, the predicted value of 8.8 eV [59] seems to be too small to compete with NC**X**, even considering the lower level of theory used for that prediction. Changing $CF_3$ groups to cyano groups does not rise the electron affinity but provides nitrogen atoms as safe protonation sites. However, PBE0 functional (without ZPE correction) measures the proton affinity of $CB_{11}(CN)_{12}^-$ as high as 10.366 eV, while for NC**X**⁻ the same level of theory gives 9.681 eV, so the novel superacid is again beyond reach.

The last idea was to use our superhalogen itself as an electron-withdrawing group, a concept applied in other form in hyperhalogens. We took boron as a central atom extending the analogy to hydrofluoric (moderately strong) and tetrafluoroboric (very strong) acids. While B—N bonds are strong enough to make boron nitride almost as hard as diamond, we can expect stability of the "tetrasuperhalogenoborate" anion B[NC**X**]$_4^-$ (Figure 10). Its ground state appeared to be a singlet of *T* symmetry with surprisingly short B—N bonds of 1.517Å.

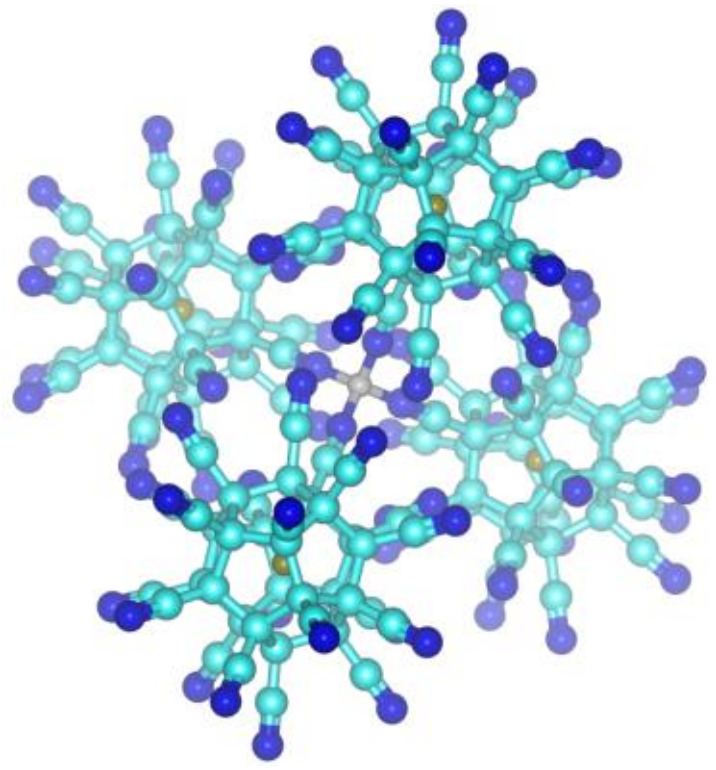

Figure 10. Structure of B[NC**X**]$_4^-$ anion.

Unfortunately, this structure was too big for us to handle it with DLPNO-CCSD(T), say nothing about the possible ways of decomposition. However, Hartree-Fock HOMO energy for B[NC**X**]$_4^-$ is 1.21 eV lower than for NC**X**⁻, providing the hope for record-breaking electron affinity around 12 eV for B[NC**X**]$_4$. The latter molecule in its ground state seems to have $D_2$ symmetry, although we did not calculate harmonic frequencies for that structure to ensure this. As for proton affinity, there's little chance for B[NC**X**]$_4^-$ to make a new record because of bridge-like protonation: the proton in HB[NC**X**]$_4$ sticks between two CN needles of adjoining cages.

## Conclusions

We designed a superhalogen F@$C_{20}(CN)_{20}$, named NC**X**, with electron affinity of 10.8 eV, and considered some of its derivatives: a fluorine superhalogenide FNC**X**, a superhalogen dimer [NC**X**]$_2$, a hyperhalogen B[NC**X**]$_4$ with electron affinity close to 12 eV, and a superacid HNC**X** with gas-phase deprotonation energy of just 208 kcal/mol.

Being much stronger acid than the current Brønsted champion, fluorocarborane acid, HNC**X** is expected to have a high degree of autoionization in condensed phase, and to be close to protonation of molecular nitrogen. $SbF_5$ is suggested as a possible solvent for HNC**X** to attest its strength.

Both HNC**X** and NC**X**⁻ anion were predicted to be thermodynamically stable, while neutral NC**X** radical is expected to have only kinetic stability at low temperatures.

Various ways of possible improvements to superhalogenic structures were investigated. The instability of some earlier-claimed superacids of record strength was established.

## Acknowledgments

D. I. S. would like to thank Deutsche Forschungsgemeinschaft (DFG-SFB 953 "Synthetic Carbon Allotropes"). The manuscript was written through contributions of both authors. Both authors have given approval to the final version of the manuscript. Both authors contributed equally (computations by D.I.S., design by A.V.K.).

**Keywords:**

Superhalogen, superacid, DLPNO-CCSD(T)

Additional Supporting Information (XYZ coordinates and energies for discussed structures) may be found in the online version of this article.

**Supporting Information**

Energy and geometry data is provided for the following structures:

**1.** Superhalogen NC**X**
**2.** Superhalogen anion NC**X**$^-$
**3.** Superacid HNC**X**
**4.** Protonated superacid HNC**X**H$^+$
**5.** Superacid bridged anion H[NC**X**]$_2^-$
**6.** Fluorocarborane acid $HCHB_{11}F_{11}$
**7.** Fluorocarborane anion $CHB_{11}F_{11}^-$
**8.** Fluorocarborane bridged anion $H[CHB_{11}F_{11}]_2^-$
**9.** Nitrogen NN
**10.** Ion pair [NNHNN]$^+$[NC**X**]$^-$
**11.** Acetonitrile $CH_3CN$
**12.** 1,2-dicyanocyclopentadiene $HC_5H_3(CN)_2$
**13.** Pentacyanotoluene $HCH_2C_6(CN)_5$
**14.** Ion pair $[(CH_3CN)_2H]^+[C_5H_3(CN)_2]^-$
**15.** Ion pair $[(CH_3CN)_2H]^+[CH_2C_6(CN)_5]^-$
**16.** Fluorine superhalogenide FNC**X**
**17.** Superhalogen dimer [NC**X**]$_2$
**18.** Empty cage $C_{20}(CN)_{20}$
**19.** Neon complex $Ne@C_{20}(CN)_{20}$
**20.** Concaved cage **X**
**21.** Destructed superhalogen
**22.** Superhalogen transition state
**23.** Superacid transition state
**24.** Alternative superhalogen $F@C_{24}(CN)_{24}$
**25.** Alternative superhalogen anion $F@C_{24}(CN)_{24}^-$
**26.** Alternative superhalogen $F@C_{28}(CN)_{28}$
**27.** Alternative superhalogen anion $F@C_{28}(CN)_{28}^-$
**28.** Cyanocarborane acid $HB_{11}C(CN)_{12}$
**29.** Cyanocarborane anion $B_{11}C(CN)_{12}^-$
**30.** Hyperhalogen B[NC**X**]$_4$
**31.** Hyperhalogen anion B[NC**X**]$_4^-$

**1**. Superhalogen NC**X**

| | |
|---|---|
| Stoichiometry: | $C_{40}N_{20}F$ |
| Charge: | 0 |
| Multiplicity: | 2 |
| Point group: | $D_{3d}$ |

Energy calculations (Hartree), cc-pVTZ, maug-cc-pVTZ on N

| | |
|---|---|
| PBE0: | -2716.0100217 |
| PBE0 ZPE correction: | 0.3092484 |
| PBE0 thermal Gibbs correction: | 0.231783 |
| DLPNO-CCSD, UHF/TightPNO: | -2713.1960624 |
| DLPNO-CCSD(T), UHF/TightPNO: | -2713.7805136 |

PBE0 entropy (cal/mol-Kelvin): 269.365

Nuclear coordinates (Angströms)

```
F   0.000000   0.000000   0.000000
C   0.000000  -1.479589   1.656791
C   0.000000   0.000000   2.225960
C   0.000000   0.000000  -2.225960
C   0.000000   1.479589  -1.656791
C   0.790966  -1.934109  -0.739823
C  -0.790966  -1.934109  -0.739823
C   0.790966   1.934109   0.739823
C  -0.790966   1.934109   0.739823
C   2.070471  -0.282058   0.739823
C  -2.070471  -0.282058   0.739823
C   2.070471   0.282058  -0.739823
C  -2.070471   0.282058  -0.739823
C   1.279505  -1.652052   0.739823
C  -1.279505  -1.652052   0.739823
C   1.281362   0.739794   1.656791
C  -1.281362   0.739794   1.656791
C   1.281362  -0.739794  -1.656791
C  -1.281362  -0.739794  -1.656791
C   1.279505   1.652052  -0.739823
C  -1.279505   1.652052  -0.739823
C   3.438821  -0.463624   1.214727
N   4.523769  -0.603011   1.577058
C   0.000000  -2.462170   2.735100
N   0.000000  -3.248551   3.574861
C   0.000000   0.000000   3.683923
N   0.000000   0.000000   4.832469
C   0.000000   0.000000  -3.683923
N   0.000000   0.000000  -4.832469
C   0.000000   2.462170  -2.735100
N   0.000000   3.248551  -3.574861
C   1.317900  -3.209919  -1.214727
N   1.739662  -4.219205  -1.577058
C  -1.317900  -3.209919  -1.214727
N  -1.739662  -4.219205  -1.577058
C   1.317900   3.209919   1.214727
N   1.739662   4.219205   1.577058
C  -1.317900   3.209919   1.214727
N  -1.739662   4.219205   1.577058
C  -3.438821  -0.463624   1.214727
N  -4.523769  -0.603011   1.577058
C   3.438821   0.463624  -1.214727
N   4.523769   0.603011  -1.577058
C  -3.438821   0.463624  -1.214727
N  -4.523769   0.603011  -1.577058
C   2.120921  -2.746294   1.214727
N   2.784107  -3.616194   1.577058
C  -2.120921  -2.746294   1.214727
N  -2.784107  -3.616194   1.577058
C   2.132302   1.231085   2.735100
N   2.813328   1.624276   3.574861
C  -2.132302   1.231085   2.735100
N  -2.813328   1.624276   3.574861
C   2.132302  -1.231085  -2.735100
N   2.813328  -1.624276  -3.574861
C  -2.132302  -1.231085  -2.735100
N  -2.813328  -1.624276  -3.574861
C   2.120921   2.746294  -1.214727
N   2.784107   3.616194  -1.577058
C  -2.120921   2.746294  -1.214727
N  -2.784107   3.616194  -1.577058
```

**2.** Superhalogen anion NC**X**$^{-}$

| | |
|---|---|
| Stoichiometry: | $C_{40}N_{20}F$ |
| Charge: | −1 |
| Multiplicity: | 1 |
| Point group: | $I_h$ |

| Energy calculations (Hartree), cc-pVTZ, maug-cc-pVTZ on N | |
|---|---|
| PBE0: | -2716.3689041 |
| PBE0 ZPE correction: | 0.3151805 |
| PBE0 thermal Gibbs correction: | 0.242493 |
| DLPNO-CCSD, RHF/NormalPNO: | -2713.6443332 |
| DLPNO-CCSD(T), RHF/NormalPNO: | -2714.1930136 |
| DLPNO-CCSD, UHF/TightPNO: | -2713.6160281 |
| DLPNO-CCSD(T), UHF/TightPNO: | -2714.1815628 |

PBE0 entropy (cal/mol-Kelvin): 255.820

Nuclear coordinates (angströms)

```
F   0.000000   0.000000   0.000000
C  -1.282753  -0.416792   1.765558
C  -0.792785   1.091175   1.765558
C   0.792785  -1.091175  -1.765558
C   1.282753   0.416792  -1.765558
C   0.000000  -2.182350   0.416792
C  -1.282753  -1.765558  -0.416792
C   1.282753   1.765558   0.416792
C   0.000000   2.182350  -0.416792
C   1.282753  -0.416792   1.765558
C  -2.075538   0.674383  -0.416792
C   2.075538  -0.674383   0.416792
C  -1.282753   0.416792  -1.765558
C   0.000000  -1.348766   1.765558
C  -2.075538  -0.674383   0.416792
C   0.792785   1.091175   1.765558
C  -1.282753   1.765558   0.416792
C   1.282753  -1.765558  -0.416792
C  -0.792785  -1.091175  -1.765558
C   2.075538   0.674383  -0.416792
C   0.000000   1.348766  -1.765558
C   2.125553  -0.690634   2.925573
N   2.788535  -0.906050   3.838089
C  -2.125553  -0.690634   2.925573
N  -2.788535  -0.906050   3.838089
C  -1.313664   1.808103   2.925573
N  -1.723409   2.372069   3.838089
C   1.313664  -1.808103  -2.925573
N   1.723409  -2.372069  -3.838089
C   2.125553   0.690634  -2.925573
N   2.788535   0.906050  -3.838089
C   0.000000  -3.616207   0.690634
N   0.000000  -4.744139   0.906050
C  -2.125553  -2.925573  -0.690634
N  -2.788535  -3.838089  -0.906050
C   2.125553   2.925573   0.690634
N   2.788535   3.838089   0.906050
C   0.000000   3.616207  -0.690634
N   0.000000   4.744139  -0.906050
C  -3.439217   1.117469  -0.690634
N  -4.511944   1.466020  -0.906050
C   3.439217  -1.117469   0.690634
N   4.511944  -1.466020   0.906050
C  -2.125553   0.690634  -2.925573
N  -2.788535   0.906050  -3.838089
C   0.000000  -2.234939   2.925573
N   0.000000  -2.932039   3.838089
C  -3.439217  -1.117469   0.690634
N  -4.511944  -1.466020   0.906050
C   1.313664   1.808103   2.925573
N   1.723409   2.372069   3.838089
C  -2.125553   2.925573   0.690634
N  -2.788535   3.838089   0.906050
C   2.125553  -2.925573  -0.690634
N   2.788535  -3.838089  -0.906050
C  -1.313664  -1.808103  -2.925573
N  -1.723409  -2.372069  -3.838089
C   3.439217   1.117469  -0.690634
N   4.511944   1.466020  -0.906050
C   0.000000   2.234939  -2.925573
N   0.000000   2.932039  -3.838089
```

**3.** Superacid HNC**X**

```
Stoichiometry:      HC40N20F
Charge:             0
Multiplicity:       1
Point group:        C3v
```

Stoichiometry: $HC_{40}N_{20}F$
Charge: 0
Multiplicity: 1
Point group: $C_{3v}$

Energy calculations (Hartree), cc-pVTZ, maug-cc-pVTZ on N

| | |
|---|---|
| PBE0: | -2716.7246726 |
| PBE0 ZPE correction: | 0.3253410 |
| PBE0 thermal Gibbs correction: | 0.252933 |
| DLPNO-CCSD, RHF/NormalPNO: | -2713.9966111 |
| DLPNO-CCSD(T), RHF/NormalPNO: | -2714.5450357 |

PBE0 entropy (cal/mol-Kelvin): 256.707

Nuclear coordinates (angströms)

```
H   0.000000   0.000000  -5.757819
F   0.000000   0.000000   0.015827
C   1.289871   0.744707  -1.607853
C   0.793709   1.945020  -0.700487
C  -0.793710  -1.941380   0.779238
C  -1.283504  -0.741031   1.693245
C   0.000000  -1.489415  -1.607853
C   1.287582  -1.659882  -0.700487
C  -1.284430   1.658063   0.779238
C   0.000000   1.482063   1.693245
C  -1.289871   0.744707  -1.607853
C   2.078140   0.283317   0.779238
C  -2.081291  -0.285138  -0.700487
C   1.283504  -0.741031   1.693245
C   0.000000   0.000000  -2.146199
C   2.081291  -0.285138  -0.700487
C  -0.793709   1.945020  -0.700487
C   1.284430   1.658063   0.779238
C  -1.287582  -1.659882  -0.700487
C   0.793710  -1.941380   0.779238
C  -2.078140   0.283317   0.779238
C   0.000000   0.000000   2.257950
C  -2.090363   1.206872  -2.739580
N  -2.629405   1.518088  -3.705526
C   2.090363   1.206872  -2.739580
N   2.629405   1.518088  -3.705526
C   1.318949   3.206667  -1.209988
N   1.735549   4.182983  -1.648778
C  -1.316711  -3.216394   1.255253
N  -1.732392  -4.225169   1.613703
C  -2.127728  -1.228444   2.777239
N  -2.797689  -1.615246   3.625980
C   0.000000  -2.413744  -2.739580
N   0.000000  -3.036176  -3.705526
C   2.117580  -2.745577  -1.209988
N   2.754795  -3.594521  -1.648778
C  -2.127123   2.748502   1.255253
N  -2.792908   3.612880   1.613703
C   0.000000   2.456889   2.777239
N   0.000000   3.230493   3.625980
C   3.443834   0.467892   1.255253
N   4.525300   0.612289   1.613703
C  -3.436529  -0.461090  -1.209988
N  -4.490344  -0.588462  -1.648778
C   2.127728  -1.228444   2.777239
N   2.797689  -1.615246   3.625980
C   0.000000   0.000000  -3.614857
N   0.000000   0.000000  -4.749346
C   3.436529  -0.461090  -1.209988
N   4.490344  -0.588462  -1.648778
C  -1.318949   3.206667  -1.209988
N  -1.735549   4.182983  -1.648778
C   2.127123   2.748502   1.255253
N   2.792908   3.612880   1.613703
C  -2.117580  -2.745577  -1.209988
N  -2.754795  -3.594521  -1.648778
C   1.316711  -3.216394   1.255253
N   1.732392  -4.225169   1.613703
C  -3.443834   0.467892   1.255253
N  -4.525300   0.612289   1.613703
C   0.000000   0.000000   3.715762
N   0.000000   0.000000   4.864140
```

**4**. Protonated superacid HNC**X**H$^+$

| | |
|---|---|
| Stoichiometry: | $H_2C_{40}N_{20}F$ |
| Charge: | 1 |
| Multiplicity: | 1 |
| Point group: | $D_{3d}$ |

Energy calculations (Hartree), cc-pVTZ, maug-cc-pVTZ on N

| | |
|---|---|
| PBE0: | -2717.0090260 |
| PBE0 ZPE correction: | 0.3357630 |
| PBE0 thermal Gibbs correction: | 0.263424 |
| DLPNO-CCSD, RHF/NormalPNO: | -2714.2793178 |
| DLPNO-CCSD(T), RHF/NormalPNO: | -2714.8272020 |

PBE0 entropy (cal/mol-Kelvin): 257.746

Nuclear coordinates (angströms)

```
H   0.000000   0.000000   5.803343
H   0.000000   0.000000  -5.803343
F   0.000000   0.000000   0.000000
C   0.000000   0.000000   2.186761
C   0.000000  -1.489646   1.646578
C   0.000000   1.489646  -1.646578
C   0.000000   0.000000  -2.186761
C  -0.794340   1.947263   0.739197
C   0.794340   1.947263   0.739197
C  -0.794340  -1.947263  -0.739197
C   0.794340  -1.947263  -0.739197
C  -2.083549  -0.285713   0.739197
C   2.083549  -0.285713   0.739197
C  -2.083549   0.285713  -0.739197
C   2.083549   0.285713  -0.739197
C  -1.290071   0.744823   1.646578
C   1.290071   0.744823   1.646578
C  -1.289209  -1.661550   0.739197
C   1.289209  -1.661550   0.739197
C  -1.289209   1.661550  -0.739197
C   1.289209   1.661550  -0.739197
C  -1.290071  -0.744823  -1.646578
C   1.290071  -0.744823  -1.646578
C  -3.441217  -0.460365   1.237836
N  -4.505161  -0.586036   1.652536
C   0.000000   0.000000   3.657810
N   0.000000   0.000000   4.791736
C   0.000000  -2.417332   2.773077
N   0.000000  -3.050185   3.732310
C   0.000000   2.417332  -2.773077
N   0.000000   3.050185  -3.732310
C   0.000000   0.000000  -3.657810
N   0.000000   0.000000  -4.791736
C  -1.321921   3.210364   1.237836
N  -1.745058   4.194602   1.652536
C   1.321921   3.210364   1.237836
N   1.745058   4.194602   1.652536
C  -1.321921  -3.210364  -1.237836
N  -1.745058  -4.194602  -1.652536
C   1.321921  -3.210364  -1.237836
N   1.745058  -4.194602  -1.652536
C   3.441217  -0.460365   1.237836
N   4.505161  -0.586036   1.652536
C  -3.441217   0.460365  -1.237836
N  -4.505161   0.586036  -1.652536
C   3.441217   0.460365  -1.237836
N   4.505161   0.586036  -1.652536
C  -2.093471   1.208666   2.773077
N  -2.641538   1.525092   3.732310
C   2.093471   1.208666   2.773077
N   2.641538   1.525092   3.732310
C  -2.119296  -2.749999   1.237836
N  -2.760103  -3.608566   1.652536
C   2.119296  -2.749999   1.237836
N   2.760103  -3.608566   1.652536
C  -2.119296   2.749999  -1.237836
N  -2.760103   3.608566  -1.652536
C   2.119296   2.749999  -1.237836
N   2.760103   3.608566  -1.652536
C  -2.093471  -1.208666  -2.773077
N  -2.641538  -1.525092  -3.732310
C   2.093471  -1.208666  -2.773077
N   2.641538  -1.525092  -3.732310
```

**5.** Superacid bridged anion H[NC**X**]$_2^-$

Stoichiometry: $HC_{80}N_{40}F_2$

Charge: −1

Multiplicity: 1

Point group: $D_{3d}$

Energy calculations (Hartree), cc-pVTZ, maug-cc-pVTZ on N

| | |
|---|---|
| PBE0: | -5433.1409233 |
| PBE0 ZPE correction: | 0.6376640 |
| PBE0 thermal Gibbs correction: | 0.509806 |
| DLPNO-CCSD, RHF/NormalPNO: | -5427.6881324 |
| DLPNO-CCSD(T), RHF/NormalPNO: | -5428.7861835 |

PBE0 entropy (cal/mol-Kelvin): 478.922

Nuclear coordinates (angströms)

```
H   0.000000   0.000000   0.000000
F   0.000000   0.000000   6.054358
C   1.286592  -0.742814   4.415022
C   2.078537   0.284230   5.326475
C  -2.076770  -0.283107   6.806804
C  -1.283055   0.740772   7.721468
C  -1.286592  -0.742814   4.415022
C  -0.793119  -1.942181   5.326475
C   0.793207   1.940090   6.806804
C   1.283055   0.740772   7.721468
C   0.000000   1.485629   4.415022
C   1.283563  -1.656982   6.806804
C  -1.285419   1.657951   5.326475
C   0.000000  -1.481545   7.721468
C   0.000000   0.000000   3.864532
C   0.793119  -1.942181   5.326475
C   1.285419   1.657951   5.326475
C   2.076770  -0.283107   6.806804
C  -2.078537   0.284230   5.326475
C  -1.283563  -1.656982   6.806804
C  -0.793207   1.940090   6.806804
C   0.000000   0.000000   8.286656
C   0.000000   2.434056   3.304779
N   0.000000   3.122650   2.385355
C   2.107954  -1.217028   3.304779
N   2.704294  -1.561325   2.385355
C   3.438312   0.464689   4.829396
N   4.503310   0.600390   4.421591
C  -3.441144  -0.468240   7.289033
N  -4.517319  -0.613439   7.662485
C  -2.126156   1.227537   8.807946
N  -2.791486   1.611665   9.661431
C  -2.107954  -1.217028   3.304779
N  -2.704294  -1.561325   2.385355
C  -1.316723  -3.210011   4.829396
N  -1.731702  -4.200176   4.421591
C   1.315064   3.214238   7.289033
N   1.727405   4.218833   7.662485
C   2.126156   1.227537   8.807946
N   2.791486   1.611665   9.661431
C   2.126080  -2.745998   7.289033
N   2.789914  -3.605393   7.662485
C  -2.121589   2.745321   4.829396
N  -2.771608   3.599786   4.421591
C   0.000000  -2.455073   8.807946
N   0.000000  -3.223331   9.661431
C   0.000000   0.000000   2.397544
N   0.000000   0.000000   1.258871
C   1.316723  -3.210011   4.829396
N   1.731702  -4.200176   4.421591
C   2.121589   2.745321   4.829396
N   2.771608   3.599786   4.421591
C   3.441144  -0.468240   7.289033
N   4.517319  -0.613439   7.662485
C  -3.438312   0.464689   4.829396
N  -4.503310   0.600390   4.421591
C  -2.126080  -2.745998   7.289033
N  -2.789914  -3.605393   7.662485
C  -1.315064   3.214238   7.289033
N  -1.727405   4.218833   7.662485
C   0.000000   0.000000   9.745492
N   0.000000   0.000000  10.893811
F   0.000000   0.000000  -6.054358
C  -1.286592   0.742814  -4.415022
C  -2.078537  -0.284230  -5.326475
C   2.076770   0.283107  -6.806804
C   1.283055  -0.740772  -7.721468
C   1.286592   0.742814  -4.415022
C   0.793119   1.942181  -5.326475
C  -0.793207  -1.940090  -6.806804
C  -1.283055  -0.740772  -7.721468
C   0.000000  -1.485629  -4.415022
C  -1.283563   1.656982  -6.806804
C   1.285419  -1.657951  -5.326475
C   0.000000   1.481545  -7.721468
C   0.000000   0.000000  -3.864532
C  -0.793119   1.942181  -5.326475
C  -1.285419  -1.657951  -5.326475
C  -2.076770   0.283107  -6.806804
C   2.078537  -0.284230  -5.326475
C   1.283563   1.656982  -6.806804
C   0.793207  -1.940090  -6.806804
C   0.000000   0.000000  -8.286656
C   0.000000  -2.434056  -3.304779
N   0.000000  -3.122650  -2.385355
C  -2.107954   1.217028  -3.304779
N  -2.704294   1.561325  -2.385355
C  -3.438312  -0.464689  -4.829396
N  -4.503310  -0.600390  -4.421591
C   3.441144   0.468240  -7.289033
N   4.517319   0.613439  -7.662485
C   2.126156  -1.227537  -8.807946
N   2.791486  -1.611665  -9.661431
C   2.107954   1.217028  -3.304779
N   2.704294   1.561325  -2.385355
C   1.316723   3.210011  -4.829396
N   1.731702   4.200176  -4.421591
C  -1.315064  -3.214238  -7.289033
N  -1.727405  -4.218833  -7.662485
C  -2.126156  -1.227537  -8.807946
N  -2.791486  -1.611665  -9.661431
C  -2.126080   2.745998  -7.289033
N  -2.789914   3.605393  -7.662485
C   2.121589  -2.745321  -4.829396
N   2.771608  -3.599786  -4.421591
C   0.000000   2.455073  -8.807946
N   0.000000   3.223331  -9.661431
C   0.000000   0.000000  -2.397544
N   0.000000   0.000000  -1.258871
C  -1.316723   3.210011  -4.829396
N  -1.731702   4.200176  -4.421591
C  -2.121589  -2.745321  -4.829396
N  -2.771608  -3.599786  -4.421591
C  -3.441144   0.468240  -7.289033
N  -4.517319   0.613439  -7.662485
C   3.438312  -0.464689  -4.829396
N   4.503310  -0.600390  -4.421591
C   2.126080   2.745998  -7.289033
N   2.789914   3.605393  -7.662485
C   1.315064  -3.214238  -7.289033
N   1.727405  -4.218833  -7.662485
C   0.000000   0.000000  -9.745492
N   0.000000   0.000000 -10.893811
```

**6**. Fluorocarborane acid $HCHB_{11}F_{11}$

Stoichiometry: $H_2B_{11}CF_{11}$
Charge: 0
Multiplicity: 1
Point group: $C_S$

Energy calculations (Hartree), cc-pVTZ, maug-cc-pVTZ on F

| | |
|---|---|
| PBE0: | -1410.6951684 |
| PBE0 ZPE correction: | 0.1106439 |
| PBE0 thermal Gibbs correction: | 0.066542 |
| DLPNO-CCSD, RHF/NormalPNO: | -1409.5302555 |
| DLPNO-CCSD(T), RHF/NormalPNO: | -1409.6822524 |

PBE0 entropy (cal/mol-Kelvin): 135.538

Nuclear coordinates (angströms)

```
H  -1.771157  -2.441476   0.000000
C   1.494173   0.696174   0.000000
B   0.156073   1.778411   0.000000
B   0.105912  -1.574608   0.000000
B  -1.304056  -0.555874   0.000000
B   1.347159  -0.742543   0.919036
B   1.347159  -0.742543  -0.919036
B  -1.106679   0.922362   0.931275
B  -1.106679   0.922362  -0.931275
B   0.614810   0.829675   1.469159
B   0.614810   0.829675  -1.469159
B  -0.339218  -0.644568   1.492639
B  -0.339218  -0.644568  -1.492639
H   2.478822   1.153180   0.000000
F   0.377443   3.099747   0.000000
F  -0.414300  -2.896730   0.000000
F  -2.386885  -1.621471   0.000000
F   2.404057  -1.241853   1.570174
F   2.404057  -1.241853  -1.570174
F  -2.068073   1.540663   1.638691
F  -2.068073   1.540663  -1.638691
F   1.156319   1.476001   2.509462
F   1.156319   1.476001  -2.509462
F  -0.815047  -1.331009   2.555005
F  -0.815047  -1.331009  -2.555005
```

**7.** Fluorocarborane anion $CHB_{11}F_{11}^{-}$

Stoichiometry: $HB_{11}CF_{11}$

Charge: −1

Multiplicity: 1

Point group: $C_{5v}$

Energy calculations (Hartree), cc-pVTZ, maug-cc-pVTZ on F

| | |
|---|---|
| PBE0: | -1410.3229740 |
| PBE0 ZPE correction: | 0.1006824 |
| PBE0 thermal Gibbs correction: | 0.058132 |
| DLPNO-CCSD, RHF/NormalPNO: | -1409.1611665 |
| DLPNO-CCSD(T), RHF/NormalPNO: | -1409.3117029 |

PBE0 entropy (cal/mol-Kelvin): 131.956

Nuclear coordinates (angströms)

```
C   0.000000   0.000000  -1.630282
B   0.000000  -1.525572  -0.861773
B   0.000000   1.544443   0.644395
B   0.000000   0.000000   1.573708
B   0.896709   1.234214  -0.861773
B  -0.896709   1.234214  -0.861773
B   0.907801  -1.249481   0.644395
B  -0.907801  -1.249481   0.644395
B   1.450906  -0.471428  -0.861773
B  -1.450906  -0.471428  -0.861773
B   1.468853   0.477259   0.644395
B  -1.468853   0.477259   0.644395
H   0.000000   0.000000  -2.713547
F   0.000000  -2.650450  -1.617509
F   0.000000   2.761999   1.253980
F   0.000000   0.000000   2.935554
F   1.557896   2.144259  -1.617509
F  -1.557896   2.144259  -1.617509
F   1.623462  -2.234504   1.253980
F  -1.623462  -2.234504   1.253980
F   2.520728  -0.819034  -1.617509
F  -2.520728  -0.819034  -1.617509
F   2.626817   0.853505   1.253980
F  -2.626817   0.853505   1.253980
```

**8.** Fluorocarborane bridged anion $H[CHB_{11}F_{11}]_2^-$

Stoichiometry: $H_3B_{22}C_2F_{22}$

Charge: −1

Multiplicity: 1

Point group: $C_{2h}$

Energy calculations (Hartree), cc-pVTZ, maug-cc-pVTZ on F

| | |
|---|---|
| PBE0: | -2821.0733864 |
| PBE0 ZPE correction: | 0.2110411 |
| PBE0 thermal Gibbs correction: | 0.138006 |
| DLPNO-CCSD, RHF/NormalPNO: | -2818.7512231 |
| DLPNO-CCSD(T), RHF/NormalPNO: | -2819.0529438 |

PBE0 entropy (cal/mol-Kelvin): 241.572

Nuclear coordinates (angströms)

```
H   0.000000   0.000000   0.000000
C   0.453478   5.267456   0.000000
B   1.732273   4.133357   0.000000
B  -1.644274   3.492397   0.000000
B  -0.369432   2.251544   0.000000
B  -0.942313   4.860413   0.900244
B  -0.942313   4.860413  -0.900244
B   1.069534   2.758861   0.914579
B   1.069534   2.758861  -0.914579
B   0.709983   4.413023   1.457942
B   0.709983   4.413023  -1.457942
B  -0.606538   3.213804   1.481752
B  -0.606538   3.213804  -1.481752
H   0.738074   6.313464   0.000000
F   3.008523   4.567049   0.000000
F  -2.964518   3.195503   0.000000
F  -0.754817   0.851262   0.000000
F  -1.617637   5.826875   1.554512
F  -1.617637   5.826875  -1.554512
F   1.823899   1.868065   1.604065
F   1.823899   1.868065  -1.604065
F   1.241651   5.051153   2.519736
F   1.241651   5.051153  -2.519736
F  -1.127221   2.693107   2.616944
F  -1.127221   2.693107  -2.616944
C  -0.453478  -5.267456   0.000000
B  -1.732273  -4.133357   0.000000
B   1.644274  -3.492397   0.000000
B   0.369432  -2.251544   0.000000
B   0.942313  -4.860413  -0.900244
B   0.942313  -4.860413   0.900244
B  -1.069534  -2.758861  -0.914579
B  -1.069534  -2.758861   0.914579
B  -0.709983  -4.413023  -1.457942
B  -0.709983  -4.413023   1.457942
B   0.606538  -3.213804  -1.481752
B   0.606538  -3.213804   1.481752
H  -0.738074  -6.313464   0.000000
F  -3.008523  -4.567049   0.000000
F   2.964518  -3.195503   0.000000
F   0.754817  -0.851262   0.000000
F   1.617637  -5.826875  -1.554512
F   1.617637  -5.826875   1.554512
F  -1.823899  -1.868065  -1.604065
F  -1.823899  -1.868065   1.604065
F  -1.241651  -5.051153  -2.519736
F  -1.241651  -5.051153   2.519736
F   1.127221  -2.693107  -2.616944
F   1.127221  -2.693107   2.616944
```

**9.** Nitrogen NN
Stoichiometry: $N_2$
Charge: 0
Multiplicity: 1
Point group: $D_{\infty h}$

Energy calculations (Hartree), cc-pVTZ, maug-cc-pVTZ on N
PBE0: -109.4430008
PBE0 ZPE correction: 0.0056515
PBE0 thermal Gibbs correction: -0.012772
DLPNO-CCSD, RHF/NormalPNO: -109.3571109
DLPNO-CCSD(T), RHF/NormalPNO: -109.3744331

PBE0 entropy (cal/mol-Kelvin): 45.730

Nuclear coordinates (angströms)

```
N   0.000000   0.000000  -0.545063
N   0.000000   0.000000   0.545063
```

**10.** Ion pair $[NNHNN]^+[NC\mathbf{X}]^-$

Stoichiometry: $HC_{40}N_{24}F$
Charge: 0
Multiplicity: 1
Point group: $C_S$

Energy calculations (Hartree), cc-pVTZ, maug-cc-pVTZ on N
PBE0: -2935.5713822
PBE0 ZPE correction: 0.3371506
PBE0 thermal Gibbs correction: 0.253243
DLPNO-CCSD, RHF/NormalPNO: -2932.6731340
DLPNO-CCSD(T), RHF/NormalPNO: -2933.2585654

PBE0 entropy (cal/mol-Kelvin): 293.710

Nuclear coordinates (angströms)

```
N  -0.011528   6.466656  -2.359380
N   0.017090   6.417199  -1.274424
H   0.050537   6.351215   0.000000
N   0.017090   6.417199   1.274424
N  -0.011528   6.466656   2.359380
F   0.000742  -0.487443   0.000000
C  -0.337806   1.276567   1.283273
C  -1.716444   0.668351   0.792919
C   1.715740  -1.669007  -0.793332
C   0.337050  -2.281136  -1.283376
C   1.827228   0.764140   0.000000
C   1.784484  -0.163418   1.284636
C  -1.783763  -0.836482  -1.283879
C  -1.825319  -1.766924   0.000000
C  -0.337806   1.276567  -1.283273
C  -0.447036  -1.154024   2.077155
C   0.447440   0.154852  -2.077408
C   0.337050  -2.281136   1.283376
C   0.516061   1.647723   0.000000
C   0.447440   0.154852   2.077408
C  -1.716444   0.668351  -0.792919
C  -1.783763  -0.836482   1.283879
C   1.784484  -0.163418  -1.284636
C   1.715740  -1.669007   0.793332
C  -0.447036  -1.154024  -2.077155
C  -0.514551  -2.659697   0.000000
C  -0.549808   2.472344  -2.097851
N  -0.690077   3.461712  -2.663039
C  -0.549808   2.472344   2.097851
N  -0.690077   3.461712   2.663039
C  -2.831010   1.454638   1.313170
N  -3.682793   2.109575   1.718779
C   2.844363  -2.431049  -1.315736
N   3.738224  -3.020852  -1.730393
C   0.558312  -3.449095  -2.128390
N   0.733529  -4.364932  -2.798693
C   3.007140   1.624732   0.000000
N   3.896320   2.351767   0.000000
C   2.954226   0.071746   2.124246
N   3.872506   0.282169   2.781072
C  -2.956141  -1.045513  -2.126369
N  -3.881749  -1.190201  -2.790584
C  -3.025866  -2.595116   0.000000
N  -3.976305  -3.239671   0.000000
C  -0.738398  -1.574603   3.443118
N  -0.965864  -1.889975   4.523692
C   0.733942   0.608763  -3.434540
N   0.949771   1.005913  -4.490354
C   0.558312  -3.449095   2.128390
N   0.733529  -4.364932   2.798693
C   0.839549   3.074518   0.000000
N   1.034281   4.204376   0.000000
C   0.733942   0.608763   3.434540
N   0.949771   1.005913   4.490354
C  -2.831010   1.454638  -1.313170
N  -3.682793   2.109575  -1.718779
C  -2.956141  -1.045513   2.126369
N  -3.881749  -1.190201   2.790584
C   2.954226   0.071746  -2.124246
N   3.872506   0.282169  -2.781072
C   2.844363  -2.431049   1.315736
N   3.738224  -3.020852   1.730393
C  -0.738398  -1.574603  -3.443118
N  -0.965864  -1.889975  -4.523692
C  -0.853264  -4.078227   0.000000
N  -1.121788  -5.194756   0.000000
```

**11.** Acetonitrile $CH_3CN$

| | |
|---|---|
| Stoichiometry: | $H_3C_2N$ |
| Charge: | 0 |
| Multiplicity: | 1 |
| Point group: | $C_{3v}$ |

Energy calculations (Hartree), cc-pVTZ, maug-cc-pVTZ on N

| | |
|---|---|
| PBE0: | -132.6387035 |
| PBE0 ZPE correction: | 0.0453593 |
| PBE0 thermal Gibbs correction: | 0.022397 |
| DLPNO-CCSD, RHF/NormalPNO: | -132.5036745 |
| DLPNO-CCSD(T), RHF/NormalPNO: | -132.5272271 |

PBE0 entropy (cal/mol-Kelvin): 57.893

Nuclear coordinates (angströms)

```
H   0.000000   1.024009  -1.545714
H   0.886818  -0.512005  -1.545714
H  -0.886818  -0.512005  -1.545714
C   0.000000   0.000000  -1.171141
C   0.000000   0.000000   0.278299
N   0.000000   0.000000   1.427741
```

**12.** 1,2-dicyanocyclopentadiene $HC_5H_3(CN)_2$

| | |
|---|---|
| Stoichiometry: | $H_4C_7N_2$ |
| Charge: | 0 |
| Multiplicity: | 1 |
| Point group: | $C_S$ |

Energy calculations (Hartree), cc-pVTZ, maug-cc-pVTZ on N

| | |
|---|---|
| PBE0: | -378.2663159 |
| PBE0 ZPE correction: | 0.0905533 |
| PBE0 thermal Gibbs correction: | 0.058225 |
| DLPNO-CCSD, RHF/NormalPNO: | -377.8430334 |
| DLPNO-CCSD(T), RHF/NormalPNO: | -377.9188934 |

PBE0 entropy (cal/mol-Kelvin): 85.846

Nuclear coordinates (angströms)

| | | | |
|---|---|---|---|
| C | 0.000000 | 0.667813 | 0.000000 |
| C | -0.128641 | -0.685326 | 0.000000 |
| C | -1.543365 | -1.035711 | 0.000000 |
| C | -1.365368 | 1.275839 | 0.000000 |
| C | -2.266914 | 0.093129 | 0.000000 |
| H | -1.909812 | -2.051531 | 0.000000 |
| H | -3.344846 | 0.162428 | 0.000000 |
| H | -1.516875 | 1.918997 | 0.875085 |
| H | -1.516875 | 1.918997 | -0.875085 |
| C | 1.194541 | 1.414573 | 0.000000 |
| N | 2.149566 | 2.061712 | 0.000000 |
| C | 0.927963 | -1.630789 | 0.000000 |
| N | 1.761734 | -2.425434 | 0.000000 |

**13.** Pentacyanotoluene $HCH_2C_6(CN)_5$

Stoichiometry: $H_3C_{12}N_5$
Charge: 0
Multiplicity: 1
Point group: $C_S$

Energy calculations (Hartree), cc-pVTZ, maug-cc-pVTZ on N

| | |
|---|---|
| PBE0: | -732.1203514 |
| PBE0 ZPE correction: | 0.1202144 |
| PBE0 thermal Gibbs correction: | 0.076745 |
| DLPNO-CCSD, RHF/NormalPNO: | -731.3270836 |
| DLPNO-CCSD(T), RHF/NormalPNO: | -731.4760134 |

PBE0 entropy (cal/mol-Kelvin): 125.852

Nuclear coordinates (angströms)

```
C   0.000000   1.254790   0.000000
C   1.206315   0.547978   0.000000
C  -1.207828   0.547175   0.000000
C   1.199705  -0.855560   0.000000
C  -1.196140  -0.852813   0.000000
C   0.002364  -1.578956   0.000000
C  -0.023233  -3.068860   0.000000
H  -0.560853  -3.438303   0.876363
H   0.979182  -3.490116   0.000000
H  -0.560853  -3.438303  -0.876363
C  -0.000656   2.677829   0.000000
N  -0.002196   3.828454   0.000000
C  -2.445242   1.250754   0.000000
N  -3.448371   1.814605   0.000000
C  -2.427986  -1.569395   0.000000
N  -3.402566  -2.181945   0.000000
C   2.442776  -1.554293   0.000000
N   3.437314  -2.133760   0.000000
C   2.442077   1.254421   0.000000
N   3.442908   1.822404   0.000000
```

**14.** Ion pair $[(CH_3CN)_2H]^+[C_5H_3(CN)_2]^-$

Stoichiometry: $H_{10}C_{11}N_4$

Charge: 0

Multiplicity: 1

Point group: $C_{2v}$

Energy calculations (Hartree), cc-pVTZ, maug-cc-pVTZ on N

| | |
|---|---|
| PBE0: | -643.5261260 |
| PBE0 ZPE correction: | 0.1796041 |
| PBE0 thermal Gibbs correction: | 0.129331 |
| DLPNO-CCSD, RHF/NormalPNO: | -642.8128879 |
| DLPNO-CCSD(T), RHF/NormalPNO: | -642.9393882 |

PBE0 entropy (cal/mol-Kelvin): 144.967

Nuclear coordinates (angströms)

```
C   0.000000   0.716832   2.232570
C   0.000000  -0.716832   2.232570
C   0.000000  -1.135499   3.572028
C   0.000000   1.135499   3.572028
C   0.000000   0.000000   4.384732
H   0.000000  -2.165526   3.897099
H   0.000000   0.000000   5.465199
H   0.000000   2.165526   3.897099
C   0.000000   1.545661   1.110557
N   0.000000   2.238885   0.179113
C   0.000000  -1.545661   1.110557
N   0.000000  -2.238885   0.179113
H   0.000000   3.729756  -1.365208
H  -0.893640   4.228243  -2.849853
H   0.893640   4.228243  -2.849853
C   0.000000   3.737855  -2.462803
C   0.000000   2.355823  -2.846119
N   0.000000   1.252815  -3.144718
H   0.000000   0.000000  -3.194294
N   0.000000  -1.252815  -3.144718
C   0.000000  -2.355823  -2.846119
C   0.000000  -3.737855  -2.462803
H   0.000000  -3.729756  -1.365208
H  -0.893640  -4.228243  -2.849853
H   0.893640  -4.228243  -2.849853
```

**15.** Ion pair $[(CH_3CN)_2H]^+[CH_2C_6(CN)_5]^-$

| | |
|---|---|
| Stoichiometry: | $H_9C_{16}N_7$ |
| Charge: | 0 |
| Multiplicity: | 1 |
| Point group: | $C_1$ |

Energy calculations (Hartree), cc-pVTZ, maug-cc-pVTZ on N

| | |
|---|---|
| PBE0: | -997.3624651 |
| PBE0 ZPE correction: | 0.2100321 |
| PBE0 thermal Gibbs correction: | 0.152355 |
| DLPNO-CCSD, RHF/NormalPNO: | -996.2828796 |
| DLPNO-CCSD(T), RHF/NormalPNO: | -996.4816113 |

PBE0 entropy (cal/mol-Kelvin): 175.350

Nuclear coordinates (angströms)

| | | | |
|---|---|---|---|
| H | -2.821717 | -3.613097 | -0.057776 |
| H | -3.952367 | -3.986442 | -1.394375 |
| H | -4.575517 | -3.761021 | 0.260866 |
| C | -3.826946 | -3.434616 | -0.462088 |
| C | -3.981575 | -2.026663 | -0.707387 |
| N | -4.091130 | -0.903169 | -0.895193 |
| H | -3.962089 | 0.451597 | -0.851117 |
| N | -3.774086 | 1.603154 | -0.752585 |
| C | -3.443169 | 2.669848 | -0.517464 |
| C | -3.006564 | 3.999811 | -0.216395 |
| H | -1.953941 | 3.943110 | 0.110783 |
| H | -3.090163 | 4.623993 | -1.107062 |
| H | -3.622765 | 4.408732 | 0.585667 |
| C | 1.053560 | 0.985309 | 0.443523 |
| C | 2.404100 | 0.819148 | 0.020058 |
| C | 0.295957 | -0.187256 | 0.654622 |
| C | 2.954196 | -0.424890 | -0.168450 |
| C | 0.811663 | -1.450495 | 0.446071 |
| C | 2.198690 | -1.664969 | 0.035908 |
| C | 2.731483 | -2.886419 | -0.147170 |
| H | 3.760475 | -3.004150 | -0.455861 |
| H | 2.143913 | -3.779735 | 0.014445 |
| C | 0.449152 | 2.243051 | 0.572780 |
| N | -0.106912 | 3.254729 | 0.668030 |
| C | -1.060465 | -0.050752 | 1.072797 |
| N | -2.157111 | 0.074493 | 1.402707 |
| C | -0.049162 | -2.560720 | 0.581298 |
| N | -0.769778 | -3.461704 | 0.654979 |
| C | 4.302321 | -0.540281 | -0.591103 |
| N | 5.393669 | -0.684065 | -0.936258 |
| C | 3.184794 | 1.988613 | -0.219739 |
| N | 3.786198 | 2.951520 | -0.411556 |

**16.** Fluorine superhalogenide FNC**X**

Stoichiometry: $C_{40}N_{20}F_2$
Charge: 0
Multiplicity: 1
Point group: $C_{3v}$

Energy calculations (Hartree), cc-pVTZ, maug-cc-pVTZ on N
PBE0: -2815.7842193
PBE0 ZPE correction: 0.3178725
PBE0 thermal Gibbs correction: 0.243539
DLPNO-CCSD, RHF/NormalPNO: -2812.9885468
DLPNO-CCSD(T), RHF/NormalPNO: -2813.5453373

PBE0 entropy (cal/mol-Kelvin): 262.977

Nuclear coordinates (angströms)

```
F   0.000000   0.000000   5.878777
F   0.000000   0.000000  -0.137753
C   0.000000  -1.488381   1.483640
C   0.000000   0.000000   2.018594
C   0.000000   0.000000  -2.380964
C   0.000000   1.482132  -1.816341
C   0.793815  -1.941588  -0.902470
C  -0.793815  -1.941588  -0.902470
C   0.794048   1.945193   0.577241
C  -0.794048   1.945193   0.577241
C   2.081610  -0.284930   0.577241
C  -2.081610  -0.284930   0.577241
C   2.078372   0.283330  -0.902470
C  -2.078372   0.283330  -0.902470
C   1.287562  -1.660262   0.577241
C  -1.287562  -1.660262   0.577241
C   1.288976   0.744190   1.483640
C  -1.288976   0.744190   1.483640
C   1.283564  -0.741066  -1.816341
C  -1.283564  -0.741066  -1.816341
C   1.284557   1.658258  -0.902470
C  -1.284557   1.658258  -0.902470
C   3.436267  -0.459744   1.088374
N   4.489016  -0.585121   1.530342
C   0.000000  -2.405031   2.621843
N   0.000000  -3.010423   3.598779
C   0.000000   0.000000   3.496455
N   0.000000   0.000000   4.629906
C   0.000000   0.000000  -3.838714
N   0.000000   0.000000  -4.987096
C   0.000000   2.456900  -2.900310
N   0.000000   3.230602  -3.748964
C   1.316868  -3.216566  -1.378313
N   1.732837  -4.225479  -1.736060
C  -1.316868  -3.216566  -1.378313
N  -1.732837  -4.225479  -1.736060
C   1.319983   3.205766   1.088374
N   1.737779   4.180162   1.530342
C  -1.319983   3.205766   1.088374
N  -1.737779   4.180162   1.530342
C  -3.436267  -0.459744   1.088374
N  -4.489016  -0.585121   1.530342
C   3.444062   0.467841  -1.378313
N   4.525790   0.612058  -1.736060
C  -3.444062   0.467841  -1.378313
N  -4.525790   0.612058  -1.736060
C   2.116283  -2.746022   1.088374
N   2.751238  -3.595042   1.530342
C  -2.116283  -2.746022   1.088374
N  -2.751238  -3.595042   1.530342
C   2.082818   1.202515   2.621843
N   2.607102   1.505211   3.598779
C  -2.082818   1.202515   2.621843
N  -2.607102   1.505211   3.598779
C   2.127738  -1.228450  -2.900310
N   2.797783  -1.615301  -3.748964
C  -2.127738  -1.228450  -2.900310
N  -2.797783  -1.615301  -3.748964
C   2.127193   2.748724  -1.378313
N   2.792953   3.613420  -1.736060
C  -2.127193   2.748724  -1.378313
N  -2.792953   3.613420  -1.736060
```

**17.** Superhalogen dimer [NC**X**]$_2$

Stoichiometry: $C_{80}N_{40}F_2$
Charge: 0
Multiplicity: 1
Point group: $D_{3d}$

Energy calculations (Hartree), cc-pVTZ, maug-cc-pVTZ on N
PBE0: -5432.1160581
PBE0 ZPE correction: 0.6296862
PBE0 thermal Gibbs correction: 0.502865
DLPNO-CCSD, RHF/NormalPNO: -5426.6430985
DLPNO-CCSD(T), RHF/NormalPNO: -5427.7444875

PBE0 entropy (cal/mol-Kelvin): 476.653

Nuclear coordinates (angströms)

```
F   0.000000   0.000000   5.375430
C   1.292319  -0.746121   3.763837
C   2.084285   0.285726   4.665487
C  -2.079434  -0.283412   6.145329
C  -1.283921   0.741272   7.058701
C  -1.292319  -0.746121   3.763837
C  -0.794696  -1.947907   4.665487
C   0.794275   1.942548   6.145329
C   1.283921   0.741272   7.058701
C   0.000000   1.492242   3.763837
C   1.285159  -1.659137   6.145329
C  -1.289589   1.662180   4.665487
C   0.000000  -1.482544   7.058701
C   0.000000   0.000000   3.237075
C   0.794696  -1.947907   4.665487
C   1.289589   1.662180   4.665487
C   2.079434  -0.283412   6.145329
C  -2.084285   0.285726   4.665487
C  -1.285159  -1.659137   6.145329
C  -0.794275   1.942548   6.145329
C   0.000000   0.000000   7.622991
C   0.000000   2.381286   2.601472
N   0.000000   2.923115   1.587126
C   2.062254  -1.190643   2.601472
N   2.531492  -1.461557   1.587126
C   3.436192   0.455058   4.147348
N   4.484442   0.571160   3.691790
C  -3.445438  -0.467470   6.618814
N  -4.528817  -0.611089   6.971989
C  -2.128062   1.228637   8.142062
N  -2.799109   1.616066   8.989724
C  -2.062254  -1.190643   2.601472
N  -2.531492  -1.461557   1.587126
C  -1.324004  -3.203359   4.147348
N  -1.747582  -4.169220   3.691790
C   1.317878   3.217572   6.618814
N   1.735190   4.227615   6.971989
C   2.128062   1.228637   8.142062
N   2.799109   1.616066   8.989724
C   2.127559  -2.750102   6.618814
N   2.793628  -3.616526   6.971989
C  -2.112188   2.748301   4.147348
N  -2.736860   3.598061   3.691790
C   0.000000  -2.457275   8.142062
N   0.000000  -3.232132   8.989724
C   0.000000   0.000000   1.770098
N   0.000000   0.000000   0.619073
C   1.324004  -3.203359   4.147348
N   1.747582  -4.169220   3.691790
C   2.112188   2.748301   4.147348
N   2.736860   3.598061   3.691790
C   3.445438  -0.467470   6.618814
N   4.528817  -0.611089   6.971989
C  -3.436192   0.455058   4.147348
N  -4.484442   0.571160   3.691790
C  -2.127559  -2.750102   6.618814
N  -2.793628  -3.616526   6.971989
C  -1.317878   3.217572   6.618814
N  -1.735190   4.227615   6.971989
C   0.000000   0.000000   9.080302
N   0.000000   0.000000  10.228713
F   0.000000   0.000000  -5.375430
C  -1.292319   0.746121  -3.763837
C  -2.084285  -0.285726  -4.665487
C   2.079434   0.283412  -6.145329
C   1.283921  -0.741272  -7.058701
C   1.292319   0.746121  -3.763837
C   0.794696   1.947907  -4.665487
C  -0.794275  -1.942548  -6.145329
C  -1.283921  -0.741272  -7.058701
C   0.000000  -1.492242  -3.763837
C  -1.285159   1.659137  -6.145329
C   1.289589  -1.662180  -4.665487
C   0.000000   1.482544  -7.058701
C   0.000000   0.000000  -3.237075
C  -0.794696   1.947907  -4.665487
C  -1.289589  -1.662180  -4.665487
C  -2.079434   0.283412  -6.145329
C   2.084285  -0.285726  -4.665487
C   1.285159   1.659137  -6.145329
C   0.794275  -1.942548  -6.145329
C   0.000000   0.000000  -7.622991
C   0.000000  -2.381286  -2.601472
N   0.000000  -2.923115  -1.587126
C  -2.062254   1.190643  -2.601472
N  -2.531492   1.461557  -1.587126
C  -3.436192  -0.455058  -4.147348
N  -4.484442  -0.571160  -3.691790
C   3.445438   0.467470  -6.618814
N   4.528817   0.611089  -6.971989
C   2.128062  -1.228637  -8.142062
N   2.799109  -1.616066  -8.989724
C   2.062254   1.190643  -2.601472
N   2.531492   1.461557  -1.587126
C   1.324004   3.203359  -4.147348
N   1.747582   4.169220  -3.691790
C  -1.317878  -3.217572  -6.618814
N  -1.735190  -4.227615  -6.971989
C  -2.128062  -1.228637  -8.142062
N  -2.799109  -1.616066  -8.989724
C  -2.127559   2.750102  -6.618814
N  -2.793628   3.616526  -6.971989
C   2.112188  -2.748301  -4.147348
N   2.736860  -3.598061  -3.691790
C   0.000000   2.457275  -8.142062
N   0.000000   3.232132  -8.989724
C   0.000000   0.000000  -1.770098
N   0.000000   0.000000  -0.619073
C  -1.324004   3.203359  -4.147348
N  -1.747582   4.169220  -3.691790
C  -2.112188  -2.748301  -4.147348
N  -2.736860  -3.598061  -3.691790
C  -3.445438   0.467470  -6.618814
N  -4.528817   0.611089  -6.971989
C   3.436192  -0.455058  -4.147348
N   4.484442  -0.571160  -3.691790
C   2.127559   2.750102  -6.618814
N   2.793628   3.616526  -6.971989
C   1.317878  -3.217572  -6.618814
N   1.735190  -4.227615  -6.971989
C   0.000000   0.000000  -9.080302
N   0.000000   0.000000 -10.228713
```

**18.** Empty cage $C_{20}(CN)_{20}$

Stoichiometry: $C_{40}N_{20}$
Charge: 0
Multiplicity: 1
Point group: $I_h$

Energy calculations (Hartree), cc-pVTZ, maug-cc-pVTZ on N

| | |
|---|---|
| PBE0: | -2616.3993929 |
| PBE0 ZPE correction: | 0.3119728 |
| PBE0 thermal Gibbs correction: | 0.239099 |
| DLPNO-CCSD, RHF/NormalPNO: | -2613.7167422 |
| DLPNO-CCSD(T), RHF/NormalPNO: | -2614.2570040 |

PBE0 entropy (cal/mol-Kelvin): 255.826

Nuclear coordinates (angströms)

```
C  -1.275327  -0.414379   1.755337
C  -0.788195   1.084858   1.755337
C   0.788195  -1.084858  -1.755337
C   1.275327   0.414379  -1.755337
C   0.000000  -2.169716   0.414379
C  -1.275327  -1.755337  -0.414379
C   1.275327   1.755337   0.414379
C   0.000000   2.169716  -0.414379
C   1.275327  -0.414379   1.755337
C  -2.063523   0.670479  -0.414379
C   2.063523  -0.670479   0.414379
C  -1.275327   0.414379  -1.755337
C   0.000000  -1.340958   1.755337
C  -2.063523  -0.670479   0.414379
C   0.788195   1.084858   1.755337
C  -1.275327   1.755337   0.414379
C   1.275327  -1.755337  -0.414379
C  -0.788195  -1.084858  -1.755337
C   2.063523   0.670479  -0.414379
C   0.000000   1.340958  -1.755337
C   2.114711  -0.687111   2.910650
N   2.777694  -0.902527   3.823167
C  -2.114711  -0.687111   2.910650
N  -2.777694  -0.902527   3.823167
C  -1.306963   1.798881   2.910650
N  -1.716709   2.362847   3.823167
C   1.306963  -1.798881  -2.910650
N   1.716709  -2.362847  -3.823167
C   2.114711   0.687111  -2.910650
N   2.777694   0.902527  -3.823167
C   0.000000  -3.597762   0.687111
N   0.000000  -4.725695   0.902527
C  -2.114711  -2.910650  -0.687111
N  -2.777694  -3.823167  -0.902527
C   2.114711   2.910650   0.687111
N   2.777694   3.823167   0.902527
C   0.000000   3.597762  -0.687111
N   0.000000   4.725695  -0.902527
C  -3.421675   1.111770  -0.687111
N  -4.494403   1.460320  -0.902527
C   3.421675  -1.111770   0.687111
N   4.494403  -1.460320   0.902527
C  -2.114711   0.687111  -2.910650
N  -2.777694   0.902527  -3.823167
C   0.000000  -2.223539   2.910650
N   0.000000  -2.920640   3.823167
C  -3.421675  -1.111770   0.687111
N  -4.494403  -1.460320   0.902527
C   1.306963   1.798881   2.910650
N   1.716709   2.362847   3.823167
C  -2.114711   2.910650   0.687111
N  -2.777694   3.823167   0.902527
C   2.114711  -2.910650  -0.687111
N   2.777694  -3.823167  -0.902527
C  -1.306963  -1.798881  -2.910650
N  -1.716709  -2.362847  -3.823167
C   3.421675   1.111770  -0.687111
N   4.494403   1.460320  -0.902527
C   0.000000   2.223539  -2.910650
N   0.000000   2.920640  -3.823167
```

**19.** Neon complex Ne@$C_{20}(CN)_{20}$

| | |
|---|---|
| Stoichiometry: | $C_{40}N_{20}Ne$ |
| Charge: | 0 |
| Multiplicity: | 1 |
| Point group: | $I_h$ |

Energy calculations (Hartree), cc-pVTZ, maug-cc-pVTZ on N

| | |
|---|---|
| PBE0: | -2745.1260811 |
| PBE0 ZPE correction: | 0.3131669 |
| PBE0 thermal Gibbs correction: | 0.239660 |
| DLPNO-CCSD, RHF/NormalPNO: | -2742.4011332 |
| DLPNO-CCSD(T), RHF/NormalPNO: | -2742.9484224 |

PBE0 entropy (cal/mol-Kelvin): 259.005

Nuclear coordinates (angströms)

| | | | |
|---|---|---|---|
| Ne | 0.000000 | 0.000000 | 0.000000 |
| C | -1.293541 | -0.420297 | 1.780407 |
| C | -0.799452 | 1.100352 | 1.780407 |
| C | 0.799452 | -1.100352 | -1.780407 |
| C | 1.293541 | 0.420297 | -1.780407 |
| C | 0.000000 | -2.200704 | 0.420297 |
| C | -1.293541 | -1.780407 | -0.420297 |
| C | 1.293541 | 1.780407 | 0.420297 |
| C | 0.000000 | 2.200704 | -0.420297 |
| C | 1.293541 | -0.420297 | 1.780407 |
| C | -2.092994 | 0.680055 | -0.420297 |
| C | 2.092994 | -0.680055 | 0.420297 |
| C | -1.293541 | 0.420297 | -1.780407 |
| C | 0.000000 | -1.360110 | 1.780407 |
| C | -2.092994 | -0.680055 | 0.420297 |
| C | 0.799452 | 1.100352 | 1.780407 |
| C | -1.293541 | 1.780407 | 0.420297 |
| C | 1.293541 | -1.780407 | -0.420297 |
| C | -0.799452 | -1.100352 | -1.780407 |
| C | 2.092994 | 0.680055 | -0.420297 |
| C | 0.000000 | 1.360110 | -1.780407 |
| C | 2.130879 | -0.692365 | 2.932903 |
| N | 2.793928 | -0.907802 | 3.845512 |
| C | -2.130879 | -0.692365 | 2.932903 |
| N | -2.793928 | -0.907802 | 3.845512 |
| C | -1.316956 | 1.812634 | 2.932903 |
| N | -1.726742 | 2.376657 | 3.845512 |
| C | 1.316956 | -1.812634 | -2.932903 |
| N | 1.726742 | -2.376657 | -3.845512 |
| C | 2.130879 | 0.692365 | -2.932903 |
| N | 2.793928 | 0.907802 | -3.845512 |
| C | 0.000000 | -3.625268 | 0.692365 |
| N | 0.000000 | -4.753314 | 0.907802 |
| C | -2.130879 | -2.932903 | -0.692365 |
| N | -2.793928 | -3.845512 | -0.907802 |
| C | 2.130879 | 2.932903 | 0.692365 |
| N | 2.793928 | 3.845512 | 0.907802 |
| C | 0.000000 | 3.625268 | -0.692365 |
| N | 0.000000 | 4.753314 | -0.907802 |
| C | -3.447835 | 1.120269 | -0.692365 |
| N | -4.520670 | 1.468855 | -0.907802 |
| C | 3.447835 | -1.120269 | 0.692365 |
| N | 4.520670 | -1.468855 | 0.907802 |
| C | -2.130879 | 0.692365 | -2.932903 |
| N | -2.793928 | 0.907802 | -3.845512 |
| C | 0.000000 | -2.240539 | 2.932903 |
| N | 0.000000 | -2.937710 | 3.845512 |
| C | -3.447835 | -1.120269 | 0.692365 |
| N | -4.520670 | -1.468855 | 0.907802 |
| C | 1.316956 | 1.812634 | 2.932903 |
| N | 1.726742 | 2.376657 | 3.845512 |
| C | -2.130879 | 2.932903 | 0.692365 |
| N | -2.793928 | 3.845512 | 0.907802 |
| C | 2.130879 | -2.932903 | -0.692365 |
| N | 2.793928 | -3.845512 | -0.907802 |
| C | -1.316956 | -1.812634 | -2.932903 |
| N | -1.726742 | -2.376657 | -3.845512 |
| C | 3.447835 | 1.120269 | -0.692365 |
| N | 4.520670 | 1.468855 | -0.907802 |
| C | 0.000000 | 2.240539 | -2.932903 |
| N | 0.000000 | 2.937710 | -3.845512 |

**20.** Concaved cage **X**

Stoichiometry: $C_{39}N_{19}F$
Charge: 0
Multiplicity: 1
Point group: $C_{3v}$

Energy calculations (Hartree), cc-pVTZ, maug-cc-pVTZ on N

| | |
|---|---|
| PBE0: | -2623.3494927 |
| PBE0 ZPE correction: | 0.3048446 |
| PBE0 thermal Gibbs correction: | 0.232803 |
| DLPNO-CCSD, RHF/NormalPNO: | -2620.7071718 |
| DLPNO-CCSD(T), RHF/NormalPNO: | -2621.2365236 |

PBE0 entropy (cal/mol-Kelvin): 252.365

Nuclear coordinates (angströms)

```
F   0.000000   0.000000   0.218647
C   1.281839  -0.740070   1.728891
C   0.822480  -1.993196   0.850068
C  -0.803414   1.962863  -0.632133
C  -1.289735   0.744629  -1.537491
C   0.000000   1.480141   1.728891
C   1.314918   1.708886   0.850068
C  -1.298182  -1.677208  -0.632133
C   0.000000  -1.489257  -1.537491
C  -1.281839  -0.740070   1.728891
C   2.101596  -0.285655  -0.632133
C  -2.137398   0.284310   0.850068
C   1.289735   0.744629  -1.537491
C   0.000000   0.000000   1.650010
C   2.137398   0.284310   0.850068
C  -0.822480  -1.993196   0.850068
C   1.298182  -1.677208  -0.632133
C  -1.314918   1.708886   0.850068
C   0.803414   1.962863  -0.632133
C  -2.101596  -0.285655  -0.632133
C   0.000000   0.000000  -2.097035
C  -1.783354  -1.029620   3.049909
N  -2.165079  -1.250009   4.110340
C   1.783354  -1.029620   3.049909
N   2.165079  -1.250009   4.110340
C   1.358233  -3.229214   1.377142
N   1.785738  -4.194087   1.830497
C  -1.318470   3.227872  -1.128771
N  -1.731214   4.233429  -1.499595
C  -2.122975   1.225700  -2.627293
N  -2.786356   1.608704  -3.482909
C   0.000000   2.059240   3.049909
N   0.000000   2.500018   4.110340
C   2.117465   2.790872   1.377142
N   2.739317   3.643537   1.830497
C  -2.136184  -2.755765  -1.128771
N  -2.800650  -3.615990  -1.499595
C   0.000000  -2.451400  -2.627293
N   0.000000  -3.217407  -3.482909
C   3.454654  -0.472107  -1.128771
N   4.531864  -0.617439  -1.499595
C  -3.475698   0.438342   1.377142
N  -4.525054   0.550549   1.830497
C   2.122975   1.225700  -2.627293
N   2.786356   1.608704  -3.482909
C   3.475698   0.438342   1.377142
N   4.525054   0.550549   1.830497
C  -1.358233  -3.229214   1.377142
N  -1.785738  -4.194087   1.830497
C   2.136184  -2.755765  -1.128771
N   2.800650  -3.615990  -1.499595
C  -2.117465   2.790872   1.377142
N  -2.739317   3.643537   1.830497
C   1.318470   3.227872  -1.128771
N   1.731214   4.233429  -1.499595
C  -3.454654  -0.472107  -1.128771
N  -4.531864  -0.617439  -1.499595
C   0.000000   0.000000  -3.550774
N   0.000000   0.000000  -4.699172
```

**21.** Destructed superhalogen

| | |
|---|---|
| Stoichiometry: | $C_{40}N_{20}F$ |
| Charge: | 0 |
| Multiplicity: | 2 |
| Point group: | $C_S$ |

Energy calculations (Hartree), cc-pVTZ, maug-cc-pVTZ on N

| | |
|---|---|
| PBE0: | -2716.0496002 |
| PBE0 ZPE correction: | 0.3135234 |
| PBE0 thermal Gibbs correction: | 0.237584 |

PBE0 entropy (cal/mol-Kelvin): 264.614

Nuclear coordinates (angströms)

```
F  -0.092377   0.058076   0.000000
C  -0.880155   1.265816   0.000000
C  -1.538158   0.899650   1.284598
C   1.498024  -1.019814  -1.288821
C   1.182487  -1.895287   0.000000
C  -0.195939   0.723452  -2.134629
C   0.997742   1.505868  -1.313696
C  -1.021328  -1.589454   1.296894
C   0.140736  -0.827128   2.099642
C  -2.097179  -0.524130  -0.819877
C   0.997742   1.505868   1.313696
C  -1.021328  -1.589454  -1.296894
C   2.020934   0.403362   0.801896
C  -1.538158   0.899650  -1.284598
C   0.327453   2.136201   0.000000
C  -2.097179  -0.524130   0.819877
C  -0.195939   0.723452   2.134629
C   0.140736  -0.827128  -2.099642
C   2.020934   0.403362  -0.801896
C  -0.368251  -2.243227   0.000000
C   1.498024  -1.019814   1.288821
C  -3.419851  -0.768694  -1.353767
N  -4.474442  -0.930092  -1.779175
C  -1.125609   4.251885   0.000000
N  -2.173327   4.727738   0.000000
C  -2.518763   1.803546   1.833356
N  -3.302607   2.511374   2.284079
C   2.497178  -1.668205  -2.122556
N   3.285742  -2.174701  -2.786220
C   1.979735  -3.111142   0.000000
N   2.609683  -4.071331   0.000000
C  -0.357082   1.244493  -3.475002
N  -0.515424   1.680889  -4.525607
C   1.623359   2.504057  -2.153435
N   2.091884   3.301268  -2.835122
C  -1.650173  -2.596732   2.135129
N  -2.163740  -3.379702   2.800096
C   0.257429  -1.344571   3.452962
N   0.339988  -1.734372   4.530098
C   1.623359   2.504057   2.153435
N   2.091884   3.301268   2.835122
C  -1.650173  -2.596732  -2.135129
N  -2.163740  -3.379702  -2.800096
C   3.353021   0.677938   1.315215
N   4.399294   0.914238   1.725648
C  -2.518763   1.803546  -1.833356
N  -3.302607   2.511374  -2.284079
C   0.175428   3.648332   0.000000
N   1.196914   4.379727   0.000000
C  -3.419851  -0.768694   1.353767
N  -4.474442  -0.930092   1.779175
C  -0.357082   1.244493   3.475002
N  -0.515424   1.680889   4.525607
C   0.257429  -1.344571  -3.452962
N   0.339988  -1.734372  -4.530098
C   3.353021   0.677938  -1.315215
N   4.399294   0.914238  -1.725648
C  -0.575406  -3.682322   0.000000
N  -0.746393  -4.817920   0.000000
C   2.497178  -1.668205   2.122556
N   3.285742  -2.174701   2.786220
```

**22.** Superhalogen transition state

Stoichiometry: $C_{40}N_{20}F$
Charge: 0
Multiplicity: 2
Point group: $C_S$

Energy calculations (Hartree), cc-pVTZ, maug-cc-pVTZ on N
PBE0: -2715.9777681
PBE0 ZPE correction: 0.3101156
PBE0 thermal Gibbs correction: 0.236039

PBE0 entropy (cal/mol-Kelvin): 259.901

Nuclear coordinates (angströms)

```
F  -0.011757   0.021374   0.000000
C  -0.553031   1.924230   0.000000
C  -1.168795   1.363262   1.299457
C   1.161059  -1.418917  -1.285901
C   0.596826  -2.159227   0.000000
C   0.083078   0.767967  -2.096738
C   1.421783   1.131080  -1.287302
C  -1.398853  -1.182883   1.289456
C  -0.070448  -0.810471   2.084989
C  -2.085915   0.159606  -0.800859
C   1.421783   1.131080   1.287302
C  -1.398853  -1.182883  -1.289456
C   2.082623  -0.224429   0.796735
C  -1.168795   1.363262  -1.299457
C   0.969299   1.891829   0.000000
C  -2.085915   0.159606   0.800859
C   0.083078   0.767967   2.096738
C  -0.070448  -0.810471  -2.084989
C   2.082623  -0.224429  -0.796735
C  -0.982135  -2.012190   0.000000
C   1.161059  -1.418917   1.285901
C  -3.435910   0.302149  -1.324751
N  -4.497310   0.441325  -1.741032
C  -0.465964   3.655586   0.000000
N  -1.312129   4.481562   0.000000
C  -1.908451   2.350145   2.065106
N  -2.475415   3.161189   2.648579
C   1.917120  -2.336066  -2.128241
N   2.518718  -3.050262  -2.796734
C   0.988269  -3.562374   0.000000
N   1.298338  -4.668120   0.000000
C   0.137799   1.310838  -3.445306
N   0.185473   1.776242  -4.494337
C   2.334847   1.956525  -2.062090
N   3.035252   2.676076  -2.620497
C  -2.310114  -1.943789   2.133675
N  -3.034915  -2.531503   2.803188
C  -0.117868  -1.319955   3.448870
N  -0.155300  -1.701516   4.531485
C   2.334847   1.956525   2.062090
N   3.035252   2.676076   2.620497
C  -2.310114  -1.943789  -2.133675
N  -3.034915  -2.531503  -2.803188
C   3.435807  -0.342052   1.321708
N   4.503110  -0.405892   1.741097
C  -1.908451   2.350145  -2.065106
N  -2.475415   3.161189  -2.648579
C   1.083673   3.400977   0.000000
N   1.869845   4.333044   0.000000
C  -3.435910   0.302149   1.324751
N  -4.497310   0.441325   1.741032
C   0.137799   1.310838   3.445306
N   0.185473   1.776242   4.494337
C  -0.117868  -1.319955  -3.448870
N  -0.155300  -1.701516  -4.531485
C   3.435807  -0.342052  -1.321708
N   4.503110  -0.405892  -1.741097
C  -1.628052  -3.318002   0.000000
N  -2.141089  -4.345439   0.000000
C   1.917120  -2.336066   2.128241
N   2.518718  -3.050262   2.796734
```

**23.** Superacid transition state

Stoichiometry: $HC_{40}N_{20}F$
Charge: 0
Multiplicity: 1
Point group: $C_S$

Energy calculations (Hartree), cc-pVTZ, maug-cc-pVTZ on N

PBE0: -2716.6500940
PBE0 ZPE correction: 0.3234939
PBE0 thermal Gibbs correction: 0.250084

PBE0 entropy (cal/mol-Kelvin): 258.084

Nuclear coordinates (angströms)

```
F  -0.026744   0.017533   0.000000
C  -0.866682   1.764838   0.000000
C  -1.390017   1.129854   1.299925
C   1.384155  -1.204883  -1.285283
C   0.956165  -2.031548   0.000000
C  -0.054704   0.760705  -2.097306
C   1.203355   1.353396  -1.286457
C  -1.178534  -1.416689   1.289759
C   0.066440  -0.819940   2.084809
C  -2.088304  -0.213772  -0.801971
C   1.203355   1.353396   1.286457
C  -1.178534  -1.416689  -1.289759
C   2.085496   0.131274   0.796180
C  -1.390017   1.129854  -1.299925
C   0.629941   2.040907   0.000000
C  -2.088304  -0.213772   0.801971
C  -0.054704   0.760705   2.097306
C   0.066440  -0.819940  -2.084809
C   2.085496   0.131274  -0.796180
C  -0.624397  -2.160446   0.000000
C   1.384155  -1.204883   1.285283
C  -3.442352  -0.306823  -1.325751
N  -4.511888  -0.355240  -1.741524
C  -1.117000   3.543168   0.000000
N  -2.111309   4.163585   0.000000
C  -2.286645   1.975633   2.066061
N  -2.983900   2.662837   2.666547
C   2.287427  -1.978192  -2.126888
N   3.003329  -2.578579  -2.794604
C   1.584007  -3.345950   0.000000
N   2.080103  -4.381650   0.000000
C  -0.097731   1.301294  -3.447062
N  -0.136172   1.761995  -4.498479
C   1.958470   2.308435  -2.082994
N   2.517586   3.115132  -2.679392
C  -1.943446  -2.324710   2.133763
N  -2.555097  -3.029733   2.802917
C   0.108481  -1.332104   3.447874
N   0.137854  -1.717390   4.529389
C   1.958470   2.308435   2.082994
N   2.517586   3.115132   2.679392
C  -1.943446  -2.324710  -2.133763
N  -2.555097  -3.029733  -2.802917
C   3.439392   0.246129   1.320303
N   4.502308   0.364402   1.738896
C  -2.286645   1.975633  -2.066061
N  -2.983900   2.662837  -2.666547
C   0.485487   3.530697   0.000000
N   1.243611   4.484026   0.000000
C  -3.442352  -0.306823   1.325751
N  -4.511888  -0.355240   1.741524
C  -0.097731   1.301294   3.447062
N  -0.136172   1.761995   4.498479
C   0.108481  -1.332104  -3.447874
N   0.137854  -1.717390  -4.529389
C   3.439392   0.246129  -1.320303
N   4.502308   0.364402  -1.738896
C  -1.033780  -3.558518   0.000000
N  -1.360860  -4.659344   0.000000
C   2.287427  -1.978192   2.126888
N   3.003329  -2.578579   2.794604
H   0.943346   5.453781   0.000000
```

**24.** Alternative superhalogen F@$C_{24}(CN)_{24}$

Stoichiometry: $C_{48}N_{24}F$

Charge: 0

Multiplicity: 2

Point group: $D_{6d}$

Energy calculations (Hartree), cc-pVTZ, maug-cc-pVTZ on N

PBE0: -3239.2337414

Nuclear coordinates (angströms)

```
F   0.000000   0.000000   0.000000
C   0.000000   2.508671   0.432323
C   1.254335   2.172573  -0.432323
C  -1.254335   2.172573  -0.432323
C   0.795635   1.378081  -1.734587
C  -0.795635   1.378081  -1.734587
C  -2.172573   1.254335   0.432323
C  -2.508671   0.000000  -0.432323
C  -1.591270   0.000000  -1.734587
C  -1.378081   0.795635   1.734587
C   0.000000   1.591270   1.734587
C   1.378081   0.795635   1.734587
C  -1.378081  -0.795635   1.734587
C   0.000000  -1.591270   1.734587
C   1.378081  -0.795635   1.734587
C  -2.172573  -1.254335   0.432323
C   2.508671   0.000000  -0.432323
C   2.172573   1.254335   0.432323
C   2.172573  -1.254335   0.432323
C   1.254335  -2.172573  -0.432323
C   0.000000  -2.508671   0.432323
C  -1.254335  -2.172573  -0.432323
C   1.591270   0.000000  -1.734587
C   0.795635  -1.378081  -1.734587
C  -0.795635  -1.378081  -1.734587
C  -1.229438   2.129448  -2.920264
N  -1.565878   2.712180  -3.856727
C   2.458875   0.000000  -2.920264
N   3.131756   0.000000  -3.856727
C   1.229438   2.129448  -2.920264
N   1.565878   2.712180  -3.856727
C  -2.458875   0.000000  -2.920264
N  -3.131756   0.000000  -3.856727
C  -1.229438  -2.129448  -2.920264
N  -1.565878  -2.712180  -3.856727
C   1.229438  -2.129448  -2.920264
N   1.565878  -2.712180  -3.856727
C   2.129448  -1.229438   2.920264
N   2.712180  -1.565878   3.856727
C   2.129448   1.229438   2.920264
N   2.712180   1.565878   3.856727
C   0.000000  -2.458875   2.920264
N   0.000000  -3.131756   3.856727
C   0.000000   2.458875   2.920264
N   0.000000   3.131756   3.856727
C  -2.129448  -1.229438   2.920264
N  -2.712180  -1.565878   3.856727
C  -2.129448   1.229438   2.920264
N  -2.712180   1.565878   3.856727
C  -3.915868   0.000000  -0.817188
N  -5.023282   0.000000  -1.122098
C  -1.957934  -3.391241  -0.817188
N  -2.511641  -4.350290  -1.122098
C   0.000000   3.915868   0.817188
N   0.000000   5.023282   1.122098
C  -1.957934   3.391241  -0.817188
N  -2.511641   4.350290  -1.122098
C  -3.391241   1.957934   0.817188
N  -4.350290   2.511641   1.122098
C  -3.391241  -1.957934   0.817188
N  -4.350290  -2.511641   1.122098
C   0.000000  -3.915868   0.817188
N   0.000000  -5.023282   1.122098
C   1.957934  -3.391241  -0.817188
N   2.511641  -4.350290  -1.122098
C   3.391241  -1.957934   0.817188
N   4.350290  -2.511641   1.122098
C   3.915868   0.000000  -0.817188
N   5.023282   0.000000  -1.122098
C   3.391241   1.957934   0.817188
N   4.350290   2.511641   1.122098
C   1.957934   3.391241  -0.817188
N   2.511641   4.350290  -1.122098
```

**25.** Alternative superhalogen anion $F@C_{24}(CN)_{24}^{-}$

Stoichiometry: $C_{48}N_{24}F$
Charge: −1
Multiplicity: 1
Point group: $D_{6d}$

Energy calculations (Hartree), cc-pVTZ, maug-cc-pVTZ on N
PBE0: -3239.5829212

Nuclear coordinates (angströms)

```
F   0.000000   0.000000   0.000000
C   0.000000   2.506807   0.432897
C   1.253403   2.170958  -0.432897
C  -1.253403   2.170958  -0.432897
C   0.798093   1.382337  -1.742645
C  -0.798093   1.382337  -1.742645
C  -2.170958   1.253403   0.432897
C  -2.506807   0.000000  -0.432897
C  -1.596185   0.000000  -1.742645
C  -1.382337   0.798093   1.742645
C   0.000000   1.596185   1.742645
C   1.382337   0.798093   1.742645
C  -1.382337  -0.798093   1.742645
C   0.000000  -1.596185   1.742645
C   1.382337  -0.798093   1.742645
C  -2.170958  -1.253403   0.432897
C   2.506807   0.000000  -0.432897
C   2.170958   1.253403   0.432897
C   2.170958  -1.253403   0.432897
C   1.253403  -2.170958  -0.432897
C   0.000000  -2.506807   0.432897
C  -1.253403  -2.170958  -0.432897
C   1.596185   0.000000  -1.742645
C   0.798093  -1.382337  -1.742645
C  -0.798093  -1.382337  -1.742645
C  -1.241983   2.151178  -2.912651
N  -1.596710   2.765583  -3.815532
C   2.483966   0.000000  -2.912651
N   3.193420   0.000000  -3.815532
C   1.241983   2.151178  -2.912651
N   1.596710   2.765583  -3.815532
C  -2.483966   0.000000  -2.912651
N  -3.193420   0.000000  -3.815532
C  -1.241983  -2.151178  -2.912651
N  -1.596710  -2.765583  -3.815532
C   1.241983  -2.151178  -2.912651
N   1.596710  -2.765583  -3.815532
C   2.151178  -1.241983   2.912651
N   2.765583  -1.596710   3.815532
C   2.151178   1.241983   2.912651
N   2.765583   1.596710   3.815532
C   0.000000  -2.483966   2.912651
N   0.000000  -3.193420   3.815532
C   0.000000   2.483966   2.912651
N   0.000000   3.193420   3.815532
C  -2.151178  -1.241983   2.912651
N  -2.765583  -1.596710   3.815532
C  -2.151178   1.241983   2.912651
N  -2.765583   1.596710   3.815532
C  -3.918598   0.000000  -0.806772
N  -5.031937   0.000000  -1.087285
C  -1.959299  -3.393605  -0.806772
N  -2.515968  -4.357785  -1.087285
C   0.000000   3.918598   0.806772
N   0.000000   5.031937   1.087285
C  -1.959299   3.393605  -0.806772
N  -2.515968   4.357785  -1.087285
C  -3.393605   1.959299   0.806772
N  -4.357785   2.515968   1.087285
C  -3.393605  -1.959299   0.806772
N  -4.357785  -2.515968   1.087285
C   0.000000  -3.918598   0.806772
N   0.000000  -5.031937   1.087285
C   1.959299  -3.393605  -0.806772
N   2.515968  -4.357785  -1.087285
C   3.393605  -1.959299   0.806772
N   4.357785  -2.515968   1.087285
C   3.918598   0.000000  -0.806772
N   5.031937   0.000000  -1.087285
C   3.393605   1.959299   0.806772
N   4.357785   2.515968   1.087285
C   1.959299   3.393605  -0.806772
N   2.515968   4.357785  -1.087285
```

**26.** Alternative superhalogen F@$C_{28}(CN)_{28}$

Stoichiometry: $C_{56}N_{28}F$
Charge: 0
Multiplicity: 2
Point group: $T_d$

Energy calculations (Hartree), cc-pVTZ, maug-cc-pVTZ on N
PBE0: -3762.3782380

Nuclear coordinates (angströms)

```
F   0.000000   0.000000   0.000000
C   1.882025  -0.056479   1.882025
C   0.564343   2.507059   0.564343
C  -0.056479   1.882025   1.882025
C   0.564343   0.564343   2.507059
C   1.576960  -1.576960   1.576960
C  -0.564343  -0.564343   2.507059
C   0.056479  -1.882025   1.882025
C  -1.576960   1.576960   1.576960
C   1.882025  -1.882025   0.056479
C   2.507059  -0.564343  -0.564343
C   1.882025   0.056479  -1.882025
C   0.564343  -2.507059  -0.564343
C  -0.056479  -1.882025  -1.882025
C   0.564343  -0.564343  -2.507059
C  -0.564343  -2.507059   0.564343
C   0.056479   1.882025  -1.882025
C   1.576960   1.576960  -1.576960
C  -0.564343   2.507059  -0.564343
C  -0.564343   0.564343  -2.507059
C   1.882025   1.882025  -0.056479
C   2.507059   0.564343   0.564343
C  -1.882025  -0.056479  -1.882025
C  -1.576960  -1.576960  -1.576960
C  -1.882025   0.056479   1.882025
C  -2.507059  -0.564343   0.564343
C  -1.882025  -1.882025  -0.056479
C  -2.507059   0.564343  -0.564343
C  -1.882025   1.882025   0.056479
C   2.919852   0.008481   2.919852
C   0.876296   3.909842   0.876296
C   0.008481   2.919852   2.919852
C   0.876296   0.876296   3.909842
C   2.420315  -2.420315   2.420315
C  -0.876296  -0.876296   3.909842
C  -0.008481  -2.919852   2.919852
C  -2.420315   2.420315   2.420315
C   2.919852  -2.919852  -0.008481
C   3.909842  -0.876296  -0.876296
C   2.919852  -0.008481  -2.919852
C   0.876296  -3.909842  -0.876296
C   0.008481  -2.919852  -2.919852
C   0.876296  -0.876296  -3.909842
C  -0.876296  -3.909842   0.876296
C  -0.008481   2.919852  -2.919852
C   2.420315   2.420315  -2.420315
C  -0.876296   3.909842  -0.876296
C  -0.876296   0.876296  -3.909842
C   2.919852   2.919852   0.008481
C   3.909842   0.876296   0.876296
C  -2.919852   0.008481  -2.919852
C  -2.420315  -2.420315  -2.420315
C  -2.919852  -0.008481   2.919852
C  -3.909842  -0.876296   0.876296
C  -2.919852  -2.919852   0.008481
C  -3.909842   0.876296  -0.876296
C  -2.919852   2.919852  -0.008481
N   3.732922   0.050669   3.732922
N   1.114257   5.010591   1.114257
N   0.050669   3.732922   3.732922
N   1.114257   1.114257   5.010591
N   3.083439  -3.083439   3.083439
N  -1.114257  -1.114257   5.010591
N  -0.050669  -3.732922   3.732922
N  -3.083439   3.083439   3.083439
N   3.732922  -3.732922  -0.050669
N   5.010591  -1.114257  -1.114257
N   3.732922  -0.050669  -3.732922
N   1.114257  -5.010591  -1.114257
N   0.050669  -3.732922  -3.732922
N   1.114257  -1.114257  -5.010591
N  -1.114257  -5.010591   1.114257
N  -0.050669   3.732922  -3.732922
N   3.083439   3.083439  -3.083439
N  -1.114257   5.010591  -1.114257
N  -1.114257   1.114257  -5.010591
N   3.732922   3.732922   0.050669
N   5.010591   1.114257   1.114257
N  -3.732922   0.050669  -3.732922
N  -3.083439  -3.083439  -3.083439
N  -3.732922  -0.050669   3.732922
N  -5.010591  -1.114257   1.114257
N  -3.732922  -3.732922   0.050669
N  -5.010591   1.114257  -1.114257
N  -3.732922   3.732922  -0.050669
```

**27.** Alternative superhalogen anion $F@C_{28}(CN)_{28}^{-}$

Stoichiometry: $C_{56}N_{28}F$

Charge: −1

Multiplicity: 1

Point group: $T_d$

Energy calculations (Hartree), cc-pVTZ, maug-cc-pVTZ on N

PBE0: -3762.7267553

Nuclear coordinates (angströms)

```
F   0.000000   0.000000   0.000000
C   0.056499   1.884861   1.884861
C  -2.512119   0.565416   0.565416
C  -1.884861  -0.056499   1.884861
C  -0.565416   0.565416   2.512119
C   1.576809   1.576809   1.576809
C   0.565416  -0.565416   2.512119
C   1.884861   0.056499   1.884861
C  -1.576809  -1.576809   1.576809
C   1.884861   1.884861   0.056499
C   0.565416   2.512119  -0.565416
C  -0.056499   1.884861  -1.884861
C   2.512119   0.565416  -0.565416
C   1.884861  -0.056499  -1.884861
C   0.565416   0.565416  -2.512119
C   2.512119  -0.565416   0.565416
C  -1.884861   0.056499  -1.884861
C  -1.576809   1.576809  -1.576809
C  -2.512119  -0.565416  -0.565416
C  -0.565416  -0.565416  -2.512119
C  -1.884861   1.884861  -0.056499
C  -0.565416   2.512119   0.565416
C   0.056499  -1.884861  -1.884861
C   1.576809  -1.576809  -1.576809
C  -0.056499  -1.884861   1.884861
C   0.565416  -2.512119   0.565416
C   1.884861  -1.884861  -0.056499
C  -0.565416  -2.512119  -0.565416
C  -1.884861  -1.884861   0.056499
C   0.003120   2.923437   2.923437
C  -3.916221   0.874944   0.874944
C  -2.923437  -0.003120   2.923437
C  -0.874944   0.874944   3.916221
C   2.420983   2.420983   2.420983
C   0.874944  -0.874944   3.916221
C   2.923437   0.003120   2.923437
C  -2.420983  -2.420983   2.420983
C   2.923437   2.923437   0.003120
C   0.874944   3.916221  -0.874944
C  -0.003120   2.923437  -2.923437
C   3.916221   0.874944  -0.874944
C   2.923437  -0.003120  -2.923437
C   0.874944   0.874944  -3.916221
C   3.916221  -0.874944   0.874944
C  -2.923437   0.003120  -2.923437
C  -2.420983   2.420983  -2.420983
C  -3.916221  -0.874944  -0.874944
C  -0.874944  -0.874944  -3.916221
C  -2.923437   2.923437  -0.003120
C  -0.874944   3.916221   0.874944
C   0.003120  -2.923437  -2.923437
C   2.420983  -2.420983  -2.420983
C  -0.003120  -2.923437   2.923437
C   0.874944  -3.916221   0.874944
C   2.923437  -2.923437  -0.003120
C  -0.874944  -3.916221  -0.874944
C  -2.923437  -2.923437   0.003120
N  -0.017859   3.735123   3.735123
N  -5.016088   1.108056   1.108056
N  -3.735123   0.017859   3.735123
N  -1.108056   1.108056   5.016088
N   3.083909   3.083909   3.083909
N   1.108056  -1.108056   5.016088
N   3.735123  -0.017859   3.735123
N  -3.083909  -3.083909   3.083909
N   3.735123   3.735123  -0.017859
N   1.108056   5.016088  -1.108056
N   0.017859   3.735123  -3.735123
N   5.016088   1.108056  -1.108056
N   3.735123   0.017859  -3.735123
N   1.108056   1.108056  -5.016088
N   5.016088  -1.108056   1.108056
N  -3.735123  -0.017859  -3.735123
N  -3.083909   3.083909  -3.083909
N  -5.016088  -1.108056  -1.108056
N  -1.108056  -1.108056  -5.016088
N  -3.735123   3.735123   0.017859
N  -1.108056   5.016088   1.108056
N  -0.017859  -3.735123  -3.735123
N   3.083909  -3.083909  -3.083909
N   0.017859  -3.735123   3.735123
N   1.108056  -5.016088   1.108056
N   3.735123  -3.735123   0.017859
N  -1.108056  -5.016088  -1.108056
N  -3.735123  -3.735123  -0.017859
```

**28.** Cyanocarborane acid $HB_{11}C(CN)_{12}$

| | |
|---|---|
| Stoichiometry: | $HB_{11}C_{13}N_{12}$ |
| Charge: | 0 |
| Multiplicity: | 1 |
| Point group: | $C_{5v}$ |

Energy calculations (Hartree), cc-pVTZ, maug-cc-pVTZ on N

| | |
|---|---|
| PBE0: | -1425.2286610 |

Nuclear coordinates (angströms)

| | | | |
|---|---|---|---|
| H | 0.000000 | 0.000000 | 5.300376 |
| C | 0.000000 | 0.000000 | -1.579757 |
| B | 0.000000 | 1.532087 | -0.773574 |
| B | 0.000000 | -1.536770 | 0.730568 |
| B | 0.000000 | 0.000000 | 1.614400 |
| B | -0.900538 | -1.239484 | -0.773574 |
| B | 0.900538 | -1.239484 | -0.773574 |
| B | -0.903291 | 1.243273 | 0.730568 |
| B | 0.903291 | 1.243273 | 0.730568 |
| B | -1.457101 | 0.473441 | -0.773574 |
| B | 1.457101 | 0.473441 | -0.773574 |
| B | -1.461555 | -0.474888 | 0.730568 |
| B | 1.461555 | -0.474888 | 0.730568 |
| C | -2.689450 | 0.873855 | -1.567111 |
| C | 0.000000 | 0.000000 | -3.011634 |
| C | 0.000000 | 2.827855 | -1.567111 |
| C | 0.000000 | -2.868951 | 1.472944 |
| C | 0.000000 | 0.000000 | 3.156360 |
| C | -1.662172 | -2.287783 | -1.567111 |
| C | 1.662172 | -2.287783 | -1.567111 |
| C | -1.686327 | 2.321031 | 1.472944 |
| C | 1.686327 | 2.321031 | 1.472944 |
| C | 2.689450 | 0.873855 | -1.567111 |
| C | -2.728535 | -0.886555 | 1.472944 |
| C | 2.728535 | -0.886555 | 1.472944 |
| N | -3.631290 | 1.179878 | -2.152062 |
| N | 0.000000 | 0.000000 | -4.160106 |
| N | 0.000000 | 3.818165 | -2.152062 |
| N | 0.000000 | -3.833133 | 2.100596 |
| N | 0.000000 | 0.000000 | 4.294289 |
| N | -2.244261 | -3.088960 | -2.152062 |
| N | 2.244261 | -3.088960 | -2.152062 |
| N | -2.253059 | 3.101069 | 2.100596 |
| N | 2.253059 | 3.101069 | 2.100596 |
| N | 3.631290 | 1.179878 | -2.152062 |
| N | -3.645526 | -1.184503 | 2.100596 |
| N | 3.645526 | -1.184503 | 2.100596 |

**29.** Cyanocarborane anion $B_{11}C(CN)_{12}^{-}$

Stoichiometry: $B_{11}C_{13}N_{12}$
Charge: −1
Multiplicity: 1
Point group: $C_{5v}$

Energy calculations (Hartree), cc-pVTZ, maug-cc-pVTZ on N
PBE0: -1424.8477151

Nuclear coordinates (angströms)

```
C   0.000000   0.000000  -1.512330
B   0.000000   1.523426  -0.698224
B   0.000000  -1.525522   0.811819
B   0.000000   0.000000   1.750360
B  -0.895448  -1.232478  -0.698224
B   0.895448  -1.232478  -0.698224
B  -0.896679   1.234173   0.811819
B   0.896679   1.234173   0.811819
B  -1.448865   0.470765  -0.698224
B   1.448865   0.470765  -0.698224
B  -1.450858  -0.471412   0.811819
B   1.450858  -0.471412   0.811819
C  -2.679436   0.870602  -1.504557
C   0.000000   0.000000  -2.944468
C   0.000000   2.817326  -1.504557
C   0.000000  -2.908659   1.462142
C   0.000000   0.000000   3.279386
C  -1.655983  -2.279265  -1.504557
C   1.655983  -2.279265  -1.504557
C  -1.709667   2.353154   1.462142
C   1.709667   2.353154   1.462142
C   2.679436   0.870602  -1.504557
C  -2.766299  -0.898825   1.462142
C   2.766299  -0.898825   1.462142
N  -3.610929   1.173262  -2.107262
N   0.000000   0.000000  -4.092812
N   0.000000   3.796755  -2.107262
N   0.000000  -3.951759   1.946892
N   0.000000   0.000000   4.429696
N  -2.231677  -3.071639  -2.107262
N   2.231677  -3.071639  -2.107262
N  -2.322786   3.197040   1.946892
N   2.322786   3.197040   1.946892
N   3.610929   1.173262  -2.107262
N  -3.758346  -1.221161   1.946892
N   3.758346  -1.221161   1.946892
```

**30.** Hyperhalogen B[NC**X**]$_4$

Stoichiometry: $BC_{160}N_{80}F_4$

Charge: 0

Multiplicity: 2

Point group: $D_2$

Energy calculations (Hartree), cc-pVTZ, maug-cc-pVTZ on N

PBE0: -10889.3387489

Nuclear coordinates (angströms)

```
B   0.000000   0.000000   0.000000
F   3.634174   3.635153   3.634046
C   3.779903   2.639917   1.672835
C   3.945306   4.207562   1.511724
C   3.352677   3.092692   5.781490
C   3.526538   4.660463   5.620388
C   2.639376   1.673354   3.780029
C   4.206976   1.512409   3.945671
C   3.090766   5.781478   3.352397
C   4.656864   5.618704   3.528181
C   1.672445   3.779947   2.639344
C   5.779160   3.354535   3.093484
C   1.511520   3.946165   4.207104
C   5.617115   3.528626   4.659954
C   2.384139   2.384451   2.384147
C   4.907660   2.106404   2.648743
C   2.647613   4.907644   2.104798
C   5.185855   4.648141   2.393779
C   2.105238   2.649029   4.907790
C   4.647158   2.392586   5.188270
C   2.391874   5.188238   4.647208
C   4.924559   4.926964   4.926695
C   0.375601   3.782709   1.963104
N  -0.615117   3.630664   1.403047
C   3.784187   1.963565   0.376133
N   3.634635   1.403272  -0.614830
C   4.142785   4.550857   0.108597
N   4.297714   4.790219  -1.004091
C   3.158233   2.729370   7.179721
N   3.001747   2.441764   8.281115
C   3.451147   5.327502   6.914788
N   3.394708   5.855656   7.934249
C   1.963181   0.376523   3.783068
N   1.403311  -0.614269   3.630806
C   4.550014   0.108584   4.138638
N   4.788477  -1.005532   4.284315
C   2.728173   7.179725   3.155431
N   2.443633   8.281240   2.994756
C   5.328923   6.911933   3.461984
N   5.866323   7.928993   3.419381
C   7.179133   3.164504   2.732446
N   8.283476   3.016011   2.448618
C   0.107935   4.141477   4.549833
N  -1.005652   4.292476   4.787757
C   6.913366   3.463932   5.325845
N   7.937112   3.425144   5.849815
C   1.535241   1.535171   1.535103
N   0.875663   0.875552   0.875281
C   5.726840   1.092257   1.998759
N   6.366318   0.272650   1.511064
C   1.997983   5.723372   1.087677
N   1.511524   6.356758   0.262624
C   6.196291   5.307595   1.574794
N   6.992920   5.830240   0.931127
C   1.088851   1.998976   5.724071
N   0.264814   1.511421   6.358040
C   5.305623   1.568187   6.194207
N   5.827458   0.914995   6.981979
C   1.570311   6.197743   5.304084
N   0.921676   6.992723   5.821968
C   5.775541   5.771655   5.758655
N   6.456312   6.440138   6.403766
F  -3.634174   3.635153  -3.634046
C  -3.779903   2.639917  -1.672835
C  -4.907660   2.106404  -2.648743
C  -2.391874   5.188238  -4.647208
C  -3.526538   4.660463  -5.620388
C  -1.672445   3.779947  -2.639344
C  -2.647613   4.907644  -2.104798
C  -4.647158   2.392586  -5.188270
C  -5.617115   3.528626  -4.659954
C  -2.639376   1.673354  -3.780029
C  -5.185855   4.648141  -2.393779
C  -2.105238   2.649029  -4.907790
C  -4.656864   5.618704  -3.528181
C  -2.384139   2.384451  -2.384147
C  -3.945306   4.207562  -1.511724
C  -4.206976   1.512409  -3.945671
C  -5.779160   3.354535  -3.093484
C  -1.511520   3.946165  -4.207104
C  -3.090766   5.781478  -3.352397
C  -3.352677   3.092692  -5.781490
C  -4.924559   4.926964  -4.926695
C  -1.963181   0.376523  -3.783068
N  -1.403311  -0.614269  -3.630806
C  -3.784187   1.963565  -0.376133
N  -3.634635   1.403272   0.614830
C  -5.726840   1.092257  -1.998759
N  -6.366318   0.272650  -1.511064
C  -1.570311   6.197743  -5.304084
N  -0.921676   6.992723  -5.821968
C  -3.451147   5.327502  -6.914788
N  -3.394708   5.855656  -7.934249
C  -0.375601   3.782709  -1.963104
N   0.615117   3.630664  -1.403047
C  -1.997983   5.723372  -1.087677
N  -1.511524   6.356758  -0.262624
C  -5.305623   1.568187  -6.194207
N  -5.827458   0.914995  -6.981979
C  -6.913366   3.463932  -5.325845
N  -7.937112   3.425144  -5.849815
C  -6.196291   5.307595  -1.574794
N  -6.992920   5.830240  -0.931127
C  -1.088851   1.998976  -5.724071
N  -0.264814   1.511421  -6.358040
C  -5.328923   6.911933  -3.461984
N  -5.866323   7.928993  -3.419381
C  -1.535241   1.535171  -1.535103
N  -0.875663   0.875552  -0.875281
C  -4.142785   4.550857  -0.108597
N  -4.297714   4.790219   1.004091
C  -4.550014   0.108584  -4.138638
N  -4.788477  -1.005532  -4.284315
C  -7.179133   3.164504  -2.732446
N  -8.283476   3.016011  -2.448618
C  -0.107935   4.141477  -4.549833
N   1.005652   4.292476  -4.787757
C  -2.728173   7.179725  -3.155431
N  -2.443633   8.281240  -2.994756
C  -3.158233   2.729370  -7.179721
N  -3.001747   2.441764  -8.281115
C  -5.775541   5.771655  -5.758655
N  -6.456312   6.440138  -6.403766
F  -3.634174  -3.635153   3.634046
C  -3.779903  -2.639917   1.672835
C  -4.907660  -2.106404   2.648743
C  -2.391874  -5.188238   4.647208
C  -3.526538  -4.660463   5.620388
C  -1.672445  -3.779947   2.639344
C  -2.647613  -4.907644   2.104798
C  -4.647158  -2.392586   5.188270
C  -5.617115  -3.528626   4.659954
C  -2.639376  -1.673354   3.780029
C  -5.185855  -4.648141   2.393779
C  -2.105238  -2.649029   4.907790
C  -4.656864  -5.618704   3.528181
C  -2.384139  -2.384451   2.384147
C  -3.945306  -4.207562   1.511724
C  -4.206976  -1.512409   3.945671
C  -5.779160  -3.354535   3.093484
C  -1.511520  -3.946165   4.207104
C  -3.090766  -5.781478   3.352397
C  -3.352677  -3.092692   5.781490
C  -4.924559  -4.926964   4.926695
C  -1.963181  -0.376523   3.783068
N  -1.403311   0.614269   3.630806
C  -3.784187  -1.963565   0.376133
N  -3.634635  -1.403272  -0.614830
C  -5.726840  -1.092257   1.998759
N  -6.366318  -0.272650   1.511064
C  -1.570311  -6.197743   5.304084
N  -0.921676  -6.992723   5.821968
C  -3.451147  -5.327502   6.914788
N  -3.394708  -5.855656   7.934249
C  -0.375601  -3.782709   1.963104
N   0.615117  -3.630664   1.403047
C  -1.997983  -5.723372   1.087677
N  -1.511524  -6.356758   0.262624
C  -5.305623  -1.568187   6.194207
N  -5.827458  -0.914995   6.981979
C  -6.913366  -3.463932   5.325845
N  -7.937112  -3.425144   5.849815
C  -6.196291  -5.307595   1.574794
N  -6.992920  -5.830240   0.931127
C  -1.088851  -1.998976   5.724071
N  -0.264814  -1.511421   6.358040
C  -5.328923  -6.911933   3.461984
N  -5.866323  -7.928993   3.419381
C  -1.535241  -1.535171   1.535103
N  -0.875663  -0.875552   0.875281
C  -4.142785  -4.550857   0.108597
N  -4.297714  -4.790219  -1.004091
C  -4.550014  -0.108584   4.138638
N  -4.788477   1.005532   4.284315
C  -7.179133  -3.164504   2.732446
N  -8.283476  -3.016011   2.448618
C  -0.107935  -4.141477   4.549833
N   1.005652  -4.292476   4.787757
C  -2.728173  -7.179725   3.155431
N  -2.443633  -8.281240   2.994756
C  -3.158233  -2.729370   7.179721
N  -3.001747  -2.441764   8.281115
C  -5.775541  -5.771655   5.758655
N  -6.456312  -6.440138   6.403766
F   3.634174  -3.635153  -3.634046
C   3.779903  -2.639917  -1.672835
C   3.945306  -4.207562  -1.511724
C   3.352677  -3.092692  -5.781490
C   3.526538  -4.660463  -5.620388
C   2.639376  -1.673354  -3.780029
C   4.206976  -1.512409  -3.945671
C   3.090766  -5.781478  -3.352397
C   4.656864  -5.618704  -3.528181
C   1.672445  -3.779947  -2.639344
C   5.779160  -3.354535  -3.093484
C   1.511520  -3.946165  -4.207104
C   5.617115  -3.528626  -4.659954
C   2.384139  -2.384451  -2.384147
C   4.907660  -2.106404  -2.648743
C   2.647613  -4.907644  -2.104798
C   5.185855  -4.648141  -2.393779
C   2.105238  -2.649029  -4.907790
C   4.647158  -2.392586  -5.188270
C   2.391874  -5.188238  -4.647208
C   4.924559  -4.926964  -4.926695
C   0.375601  -3.782709  -1.963104
N  -0.615117  -3.630664  -1.403047
C   3.784187  -1.963565  -0.376133
N   3.634635  -1.403272   0.614830
C   4.142785  -4.550857  -0.108597
N   4.297714  -4.790219   1.004091
C   3.158233  -2.729370  -7.179721
N   3.001747  -2.441764  -8.281115
C   3.451147  -5.327502  -6.914788
N   3.394708  -5.855656  -7.934249
C   1.963181  -0.376523  -3.783068
N   1.403311   0.614269  -3.630806
C   4.550014  -0.108584  -4.138638
N   4.788477   1.005532  -4.284315
C   2.728173  -7.179725  -3.155431
N   2.443633  -8.281240  -2.994756
C   5.328923  -6.911933  -3.461984
N   5.866323  -7.928993  -3.419381
C   7.179133  -3.164504  -2.732446
N   8.283476  -3.016011  -2.448618
C   0.107935  -4.141477  -4.549833
N  -1.005652  -4.292476  -4.787757
C   6.913366  -3.463932  -5.325845
N   7.937112  -3.425144  -5.849815
C   1.535241  -1.535171  -1.535103
N   0.875663  -0.875552  -0.875281
C   5.726840  -1.092257  -1.998759
N   6.366318  -0.272650  -1.511064
C   1.997983  -5.723372  -1.087677
N   1.511524  -6.356758  -0.262624
C   6.196291  -5.307595  -1.574794
N   6.992920  -5.830240  -0.931127
C   1.088851  -1.998976  -5.724071
N   0.264814  -1.511421  -6.358040
C   5.305623  -1.568187  -6.194207
N   5.827458  -0.914995  -6.981979
C   1.570311  -6.197743  -5.304084
N   0.921676  -6.992723  -5.821968
C   5.775541  -5.771655  -5.758655
N   6.456312  -6.440138  -6.403766
```

**31.** Hyperhalogen anion B[NC**X**]$_4^-$

Stoichiometry: $BC_{160}N_{80}F_4$

Charge: −1

Multiplicity: 1

Point group: T

Energy calculations (Hartree), cc-pVTZ, maug-cc-pVTZ on N

PBE0: -10889.7265885

Nuclear coordinates (angströms)

| | | | |
|---|---|---|---|
| B | 0.000000 | 0.000000 | 0.000000 |
| F | 3.634307 | 3.634307 | 3.634307 |
| C | 1.673568 | 3.780226 | 2.639639 |
| C | 1.512790 | 3.946218 | 4.207096 |
| C | 5.781574 | 3.353425 | 3.093113 |
| C | 5.621430 | 3.528377 | 4.661116 |
| C | 3.780226 | 2.639639 | 1.673568 |
| C | 3.946218 | 4.207096 | 1.512790 |
| C | 3.353425 | 3.093113 | 5.781574 |
| C | 3.528377 | 4.661116 | 5.621430 |
| C | 2.639639 | 1.673568 | 3.780226 |
| C | 3.093113 | 5.781574 | 3.353425 |
| C | 4.207096 | 1.512790 | 3.946218 |
| C | 4.661116 | 5.621430 | 3.528377 |
| C | 2.384462 | 2.384462 | 2.384462 |
| C | 2.649303 | 4.907604 | 2.105925 |
| C | 2.105925 | 2.649303 | 4.907604 |
| C | 2.393190 | 5.188477 | 4.647877 |
| C | 4.907604 | 2.105925 | 2.649303 |
| C | 5.188477 | 4.647877 | 2.393190 |
| C | 4.647877 | 2.393190 | 5.188477 |
| C | 4.929589 | 4.929589 | 4.929589 |
| C | 1.963299 | 0.376644 | 3.783661 |
| N | 1.403923 | -0.614673 | 3.632888 |
| C | 0.376644 | 3.783661 | 1.963299 |
| N | -0.614673 | 3.632888 | 1.403923 |
| C | 0.108670 | 4.140516 | 4.549681 |
| N | -1.004712 | 4.290205 | 4.788710 |
| C | 7.178746 | 3.155778 | 2.725687 |
| N | 8.276996 | 2.994432 | 2.431401 |
| C | 6.917046 | 3.448951 | 5.325334 |
| N | 7.938001 | 3.383341 | 5.846903 |
| C | 3.783661 | 1.963299 | 0.376644 |
| N | 3.632888 | 1.403923 | -0.614673 |
| C | 4.140516 | 4.549681 | 0.108670 |
| N | 4.290205 | 4.788710 | -1.004712 |
| C | 3.155778 | 2.725687 | 7.178746 |
| N | 2.994432 | 2.431401 | 8.276996 |
| C | 3.448951 | 5.325334 | 6.917046 |
| N | 3.383341 | 5.846903 | 7.938001 |
| C | 2.725687 | 7.178746 | 3.155778 |
| N | 2.431401 | 8.276996 | 2.994432 |
| C | 4.549681 | 0.108670 | 4.140516 |
| N | 4.788710 | -1.004712 | 4.290205 |
| C | 5.325334 | 6.917046 | 3.448951 |
| N | 5.846903 | 7.938001 | 3.383341 |
| C | 1.535290 | 1.535290 | 1.535290 |
| N | 0.875585 | 0.875585 | 0.875585 |
| C | 1.999556 | 5.723813 | 1.088835 |
| N | 1.513388 | 6.358774 | 0.264898 |
| C | 1.088835 | 1.999556 | 5.723813 |
| N | 0.264898 | 1.513388 | 6.358774 |
| C | 1.566174 | 6.195527 | 5.302295 |
| N | 0.908582 | 6.984716 | 5.815633 |
| C | 5.723813 | 1.088835 | 1.999556 |
| N | 6.358774 | 0.264898 | 1.513388 |
| C | 6.195527 | 5.302295 | 1.566174 |
| N | 6.984716 | 5.815633 | 0.908582 |
| C | 5.302295 | 1.566174 | 6.195527 |
| N | 5.815633 | 0.908582 | 6.984716 |
| C | 5.771441 | 5.771441 | 5.771441 |
| N | 6.434428 | 6.434428 | 6.434428 |
| F | -3.634307 | -3.634307 | 3.634307 |
| C | -1.673568 | -3.780226 | 2.639639 |
| C | -2.649303 | -4.907604 | 2.105925 |
| C | -4.647877 | -2.393190 | 5.188477 |
| C | -5.621430 | -3.528377 | 4.661116 |
| C | -2.639639 | -1.673568 | 3.780226 |
| C | -2.105925 | -2.649303 | 4.907604 |
| C | -5.188477 | -4.647877 | 2.393190 |
| C | -4.661116 | -5.621430 | 3.528377 |
| C | -3.780226 | -2.639639 | 1.673568 |
| C | -2.393190 | -5.188477 | 4.647877 |
| C | -4.907604 | -2.105925 | 2.649303 |
| C | -3.528377 | -4.661116 | 5.621430 |
| C | -2.384462 | -2.384462 | 2.384462 |
| C | -1.512790 | -3.946218 | 4.207096 |
| C | -3.946218 | -4.207096 | 1.512790 |
| C | -3.093113 | -5.781574 | 3.353425 |
| C | -4.207096 | -1.512790 | 3.946218 |
| C | -3.353425 | -3.093113 | 5.781574 |
| C | -5.781574 | -3.353425 | 3.093113 |
| C | -4.929589 | -4.929589 | 4.929589 |
| C | -3.783661 | -1.963299 | 0.376644 |
| N | -3.632888 | -1.403923 | -0.614673 |
| C | -0.376644 | -3.783661 | 1.963299 |
| N | 0.614673 | -3.632888 | 1.403923 |
| C | -1.999556 | -5.723813 | 1.088835 |
| N | -1.513388 | -6.358774 | 0.264898 |
| C | -5.302295 | -1.566174 | 6.195527 |
| N | -5.815633 | -0.908582 | 6.984716 |
| C | -6.917046 | -3.448951 | 5.325334 |
| N | -7.938001 | -3.383341 | 5.846903 |
| C | -1.963299 | -0.376644 | 3.783661 |
| N | -1.403923 | 0.614673 | 3.632888 |
| C | -1.088835 | -1.999556 | 5.723813 |
| N | -0.264898 | -1.513388 | 6.358774 |
| C | -6.195527 | -5.302295 | 1.566174 |
| N | -6.984716 | -5.815633 | 0.908582 |
| C | -5.325334 | -6.917046 | 3.448951 |
| N | -5.846903 | -7.938001 | 3.383341 |
| C | -1.566174 | -6.195527 | 5.302295 |
| N | -0.908582 | -6.984716 | 5.815633 |
| C | -5.723813 | -1.088835 | 1.999556 |
| N | -6.358774 | -0.264898 | 1.513388 |
| C | -3.448951 | -5.325334 | 6.917046 |
| N | -3.383341 | -5.846903 | 7.938001 |
| C | -1.535290 | -1.535290 | 1.535290 |
| N | -0.875585 | -0.875585 | 0.875585 |
| C | -0.108670 | -4.140516 | 4.549681 |
| N | 1.004712 | -4.290205 | 4.788710 |
| C | -4.140516 | -4.549681 | 0.108670 |
| N | -4.290205 | -4.788710 | -1.004712 |
| C | -2.725687 | -7.178746 | 3.155778 |
| N | -2.431401 | -8.276996 | 2.994432 |
| C | -4.549681 | -0.108670 | 4.140516 |
| N | -4.788710 | 1.004712 | 4.290205 |
| C | -3.155778 | -2.725687 | 7.178746 |
| N | -2.994432 | -2.431401 | 8.276996 |
| C | -7.178746 | -3.155778 | 2.725687 |
| N | -8.276996 | -2.994432 | 2.431401 |
| C | -5.771441 | -5.771441 | 5.771441 |
| N | -6.434428 | -6.434428 | 6.434428 |
| F | 3.634307 | -3.634307 | -3.634307 |
| C | 1.673568 | -3.780226 | -2.639639 |
| C | 2.649303 | -4.907604 | -2.105925 |
| C | 4.647877 | -2.393190 | -5.188477 |
| C | 5.621430 | -3.528377 | -4.661116 |
| C | 2.639639 | -1.673568 | -3.780226 |
| C | 2.105925 | -2.649303 | -4.907604 |
| C | 5.188477 | -4.647877 | -2.393190 |
| C | 4.661116 | -5.621430 | -3.528377 |
| C | 3.780226 | -2.639639 | -1.673568 |
| C | 2.393190 | -5.188477 | -4.647877 |
| C | 4.907604 | -2.105925 | -2.649303 |
| C | 3.528377 | -4.661116 | -5.621430 |
| C | 2.384462 | -2.384462 | -2.384462 |
| C | 1.512790 | -3.946218 | -4.207096 |
| C | 3.946218 | -4.207096 | -1.512790 |
| C | 3.093113 | -5.781574 | -3.353425 |
| C | 4.207096 | -1.512790 | -3.946218 |
| C | 3.353425 | -3.093113 | -5.781574 |
| C | 5.781574 | -3.353425 | -3.093113 |
| C | 4.929589 | -4.929589 | -4.929589 |
| C | 3.783661 | -1.963299 | -0.376644 |
| N | 3.632888 | -1.403923 | 0.614673 |
| C | 0.376644 | -3.783661 | -1.963299 |
| N | -0.614673 | -3.632888 | -1.403923 |
| C | 1.999556 | -5.723813 | -1.088835 |
| N | 1.513388 | -6.358774 | -0.264898 |
| C | 5.302295 | -1.566174 | -6.195527 |
| N | 5.815633 | -0.908582 | -6.984716 |
| C | 6.917046 | -3.448951 | -5.325334 |
| N | 7.938001 | -3.383341 | -5.846903 |
| C | 1.963299 | -0.376644 | -3.783661 |
| N | 1.403923 | 0.614673 | -3.632888 |
| C | 1.088835 | -1.999556 | -5.723813 |
| N | 0.264898 | -1.513388 | -6.358774 |
| C | 6.195527 | -5.302295 | -1.566174 |
| N | 6.984716 | -5.815633 | -0.908582 |
| C | 5.325334 | -6.917046 | -3.448951 |
| N | 5.846903 | -7.938001 | -3.383341 |
| C | 1.566174 | -6.195527 | -5.302295 |
| N | 0.908582 | -6.984716 | -5.815633 |
| C | 5.723813 | -1.088835 | -1.999556 |
| N | 6.358774 | -0.264898 | -1.513388 |
| C | 3.448951 | -5.325334 | -6.917046 |
| N | 3.383341 | -5.846903 | -7.938001 |
| C | 1.535290 | -1.535290 | -1.535290 |
| N | 0.875585 | -0.875585 | -0.875585 |
| C | 0.108670 | -4.140516 | -4.549681 |
| N | -1.004712 | -4.290205 | -4.788710 |
| C | 4.140516 | -4.549681 | -0.108670 |
| N | 4.290205 | -4.788710 | 1.004712 |
| C | 2.725687 | -7.178746 | -3.155778 |
| N | 2.431401 | -8.276996 | -2.994432 |
| C | 4.549681 | -0.108670 | -4.140516 |
| N | 4.788710 | 1.004712 | -4.290205 |
| C | 3.155778 | -2.725687 | -7.178746 |
| N | 2.994432 | -2.431401 | -8.276996 |
| C | 7.178746 | -3.155778 | -2.725687 |
| N | 8.276996 | -2.994432 | -2.431401 |
| C | 5.771441 | -5.771441 | -5.771441 |
| N | 6.434428 | -6.434428 | -6.434428 |
| F | -3.634307 | 3.634307 | -3.634307 |
| C | -1.673568 | 3.780226 | -2.639639 |
| C | -1.512790 | 3.946218 | -4.207096 |
| C | -5.781574 | 3.353425 | -3.093113 |
| C | -5.621430 | 3.528377 | -4.661116 |
| C | -3.780226 | 2.639639 | -1.673568 |
| C | -3.946218 | 4.207096 | -1.512790 |
| C | -3.353425 | 3.093113 | -5.781574 |
| C | -3.528377 | 4.661116 | -5.621430 |
| C | -2.639639 | 1.673568 | -3.780226 |
| C | -3.093113 | 5.781574 | -3.353425 |
| C | -4.207096 | 1.512790 | -3.946218 |
| C | -4.661116 | 5.621430 | -3.528377 |
| C | -2.384462 | 2.384462 | -2.384462 |
| C | -2.649303 | 4.907604 | -2.105925 |
| C | -2.105925 | 2.649303 | -4.907604 |
| C | -2.393190 | 5.188477 | -4.647877 |
| C | -4.907604 | 2.105925 | -2.649303 |
| C | -5.188477 | 4.647877 | -2.393190 |
| C | -4.647877 | 2.393190 | -5.188477 |
| C | -4.929589 | 4.929589 | -4.929589 |
| C | -1.963299 | 0.376644 | -3.783661 |
| N | -1.403923 | -0.614673 | -3.632888 |
| C | -0.376644 | 3.783661 | -1.963299 |
| N | 0.614673 | 3.632888 | -1.403923 |
| C | -0.108670 | 4.140516 | -4.549681 |
| N | 1.004712 | 4.290205 | -4.788710 |
| C | -7.178746 | 3.155778 | -2.725687 |
| N | -8.276996 | 2.994432 | -2.431401 |
| C | -6.917046 | 3.448951 | -5.325334 |
| N | -7.938001 | 3.383341 | -5.846903 |
| C | -3.783661 | 1.963299 | -0.376644 |
| N | -3.632888 | 1.403923 | 0.614673 |
| C | -4.140516 | 4.549681 | -0.108670 |
| N | -4.290205 | 4.788710 | 1.004712 |
| C | -3.155778 | 2.725687 | -7.178746 |
| N | -2.994432 | 2.431401 | -8.276996 |
| C | -3.448951 | 5.325334 | -6.917046 |
| N | -3.383341 | 5.846903 | -7.938001 |
| C | -2.725687 | 7.178746 | -3.155778 |
| N | -2.431401 | 8.276996 | -2.994432 |
| C | -4.549681 | 0.108670 | -4.140516 |
| N | -4.788710 | -1.004712 | -4.290205 |
| C | -5.325334 | 6.917046 | -3.448951 |
| N | -5.846903 | 7.938001 | -3.383341 |
| C | -1.535290 | 1.535290 | -1.535290 |
| N | -0.875585 | 0.875585 | -0.875585 |
| C | -1.999556 | 5.723813 | -1.088835 |
| N | -1.513388 | 6.358774 | -0.264898 |
| C | -1.088835 | 1.999556 | -5.723813 |
| N | -0.264898 | 1.513388 | -6.358774 |
| C | -1.566174 | 6.195527 | -5.302295 |
| N | -0.908582 | 6.984716 | -5.815633 |
| C | -5.723813 | 1.088835 | -1.999556 |
| N | -6.358774 | 0.264898 | -1.513388 |
| C | -6.195527 | 5.302295 | -1.566174 |
| N | -6.984716 | 5.815633 | -0.908582 |
| C | -5.302295 | 1.566174 | -6.195527 |
| N | -5.815633 | 0.908582 | -6.984716 |
| C | -5.771441 | 5.771441 | -5.771441 |
| N | -6.434428 | 6.434428 | -6.434428 |